\documentclass[a4paper]{article}

\usepackage[letterpaper]{geometry}
\usepackage[parfill]{parskip}
\PassOptionsToPackage{numbers, compress}{natbib}
\usepackage[numbers, compress]{natbib}
\usepackage{xr-hyper,refcount}

\usepackage{graphicx}
\usepackage[utf8]{inputenc} 
\usepackage[T1]{fontenc}    
\usepackage{url}  
\usepackage{tabularx}
\usepackage{booktabs}       
\usepackage{amsfonts}       
\usepackage{nicefrac}       
\usepackage{microtype}      
\usepackage[parfill]{parskip}
\usepackage{amsmath,amsthm,amssymb,bbm}
\usepackage{mathtools}
\usepackage{cases}
\usepackage{comment}
\usepackage{subcaption}
\usepackage{color}
\usepackage{appendix}
\usepackage{xspace}
\usepackage{enumitem}
\usepackage[english]{babel}
\usepackage{authblk}
\usepackage{float}

\usepackage[colorlinks,citecolor=blue,urlcolor=blue,linkcolor=blue,linktocpage=true]{hyperref}
\usepackage{cleveref}
\crefformat{equation}{(#2#1#3)}
\crefrangeformat{equation}{(#3#1#4) to~(#5#2#6)}
\crefname{equation}{}{}
\Crefname{equation}{}{}

\crefname{definition}{\textbf{definition}}{definitions}
\Crefname{definition}{Definition}{Definitions}
\crefname{assumption}{\textbf{assumption}}{assumptions}
\Crefname{assumption}{Assumption}{Assumptions}
\definecolor{maroon}{RGB}{192,80,77}

\usepackage[utf8]{inputenc} 
\usepackage[T1]{fontenc}    
\usepackage{lmodern}
\usepackage{hyperref}       
\usepackage{url}            
\usepackage{booktabs}       
\usepackage{amsfonts}       
\usepackage{nicefrac}       
\usepackage{microtype}      
\usepackage{tikz}
\usepackage{tabularx}
\usetikzlibrary{automata,decorations.markings,positioning,arrows}

\usepackage{amsmath}
\usepackage{bbm}

\usepackage[normalem]{ulem}

\usepackage{mathtools}

\usepackage{amssymb,amsmath,amsthm}
\usepackage{soul}
\usepackage{url}            
\usepackage{xcolor}
\usepackage{xr-hyper,refcount}
\usepackage[colorlinks]{hyperref}

\usepackage{caption}
\usepackage{subcaption}
\usepackage[export]{adjustbox}%
\usepackage{booktabs}
\usepackage[suppress]{color-edits}
\addauthor[Michael]{mf}{purple}
\addauthor[Hoda]{hh}{red}
\addauthor[Nik]{nm}{blue}

\usepackage[subtle]{savetrees}

\usepackage[rightcaption]{sidecap}
\sidecaptionvpos{figure}{t}

\usepackage{enumitem}
\newcommand{\xhdr}[1]{\vspace{1mm} \noindent\textbf{#1.}}
\newcommand{\qt}[1]{\textit{``#1''}}

\begin{document}

\title{\Large The AI Incident Database \\ as an Educational Tool to Raise Awareness of AI Harms:\\ \large A Classroom Exploration of Efficacy, Limitations, \& Future Improvements}




\author{
  Michael Feffer, Nikolas Martelaro, Hoda Heidari\\
  Carnegie Mellon University\\
  \{\href{mailto:mfeffer@andrew.cmu.edu}{\nolinkurl{mfeffer}}, \href{mailto:nikmart@andrew.cmu.edu}{\nolinkurl{nikmart}},
  \href{mailto:hheidari@andrew.cmu.edu}{\nolinkurl{hheidari}}\}@andrew.cmu.edu
}

\maketitle

\begin{abstract}
Prior work has established the importance of integrating AI ethics topics into computer and data sciences curricula. We provide evidence suggesting that one of the critical objectives of AI Ethics education must be to raise \emph{awareness of AI harms}. While there are various sources to learn about such harms, The AI Incident Database (AIID) is one of the few attempts at offering a relatively comprehensive database indexing prior instances of harms or near harms stemming from the deployment of AI technologies in the real world.
This study assesses the effectiveness of AIID as an educational tool to raise awareness regarding the prevalence and severity of AI harms in socially high-stakes domains. We present findings obtained through a classroom study conducted at an R1 institution as part of a course focused on the societal and ethical considerations around  AI and ML. 
Our qualitative findings characterize students' initial perceptions of core topics in AI ethics and their desire to close the educational gap between their technical skills and their ability to think systematically about ethical and societal aspects of their work.
We find that interacting with the database helps students better understand the \emph{magnitude and severity} of AI harms and instills in them a sense of urgency around (a) designing \emph{functional and safe} AI and (b) strengthening \emph{governance and accountability} mechanisms. Finally, we compile students' feedback about the tool and our class activity into actionable recommendations for the database development team and the broader community to improve awareness of AI harms in AI ethics education.
\end{abstract}


\section{Introduction}\label{sec:introduction}

Despite ongoing efforts in academia, public and private sectors, and civil society, incidents of AI-related harms\footnote{We use the terminology of `harms' to broadly refer to \emph{infringements upon values, principles, and rights that stakeholders and impacted community members legitimately expect to be respected in a given domain.}} continue to be on the rise. Only in the past few months, we have witnessed numerous new accounts of large language models producing harmful content, including but not limited to malware and ransomware~\citep{wp2023chat-gpt}, hate speech~\citep{intercept2022}, and inappropriate counseling and companionship~\citep{vide2023mental,Byte2023}. Just recently, self-driving cars caused a multi-car pileup~\citep{cnn2023crash}, initiated a police chase~\citep{angle2023flee}, and meddled with a firefighting operation~\citep{drive2023firefighter}. Clearly, the multitude of AI ethics frameworks~\citep{floridi2022unified} and responsible AI engineering tools~\citep{morley2020initial} have done little to slow down the rate at which such incidents occur. As AI ethics researchers and educators, we cannot help but ask ourselves: where have we failed, and how can we do better?

When designing new technologies, creators often suffer from what \citet{clarke1962hazards} describes as \textit{failures of imagination} in considering the potential impacts of technology.
Boyarskaya, Olteanu, and Crawford \cite{boyarskaya2020overcoming} introduce the concept of failures of imagination to the AI community, noting that it can be quite hard for many system developers to imagine how their AI system can fail and what kinds of harmful consequences their creations may impose on society.
The failure to imagine possible harms is especially challenging when system developers have not seen similar failures before.
For example, in a 2019 survey of Machine Learning practitioners, \citet{Holstein2019improving} found respondents often reported \textit{blind spots} around possible ML fairness issues, with over half of respondents stating that they ``\textit{discovered serious issues only after deploying a system in the real world.}''
\citeauthor{boyarskaya2020overcoming} thus challenge the AI research community by asking ``\textit{what can be done to address possible failures of imagination?}'' \cite[pg. 4]{boyarskaya2020overcoming}. We start from the premise that better processes and educational tools that expose students to past harms may help fulfill practitioners' desire for more support in anticipating and mitigating harms proactively \cite{Holstein2019improving}.

While failures and unanticipated consequences \cite{tenner1997things} of new technology are to be expected, one immediate goal we should strive for is to ensure that system developers learn from past mistakes.
However, this has often not been the case, as evidenced by failures like: 
\begin{itemize}
    \item Google systems tagging Black people as gorillas in 2015 \citep{aiid:16} and Meta systems labeling videos of Black men as primates in 2021 \citep{aiid:113}
    \item Microsoft's AI chatbot learning and repeating hate speech in 2016 \citep{aiid:6} and South Korean startup Scatter Lab's AI chatbot using offensive language with LGBTQ persons and people with disabilities in 2020 \citep{aiid:106}
    \item Tesla autopilot systems crashing into stationary emergency vehicles in 2018 \citep{aiid:320,aiid:323} and colliding with a stationary road works vehicle in 2022  \citep{aiid:221}
    \item YouTube's recommender algorithm suggesting violent and inappropriate videos to children in 2017 \citep{aiid:1} and TikTok recommending dangerous `blackout challenge' videos to young users in 2021 \citep{aiid:286}
\end{itemize}
Why is it that such AI failures repeat themselves? We believe one major source is practitioners' lack of awareness around prior instances of AI harms, which in turn results in failure to imagine similar harms in new use cases. 

Prior work has proposed several tools including structured checklists~\citep{madaio2020co}, responsible AI engineering guidelines \citep{dunnmon2021responsible,googlePAIR2019} and playbooks \cite{Hong2021AIPlaybook,ms-responsibleAI2022}, and card games \cite{ballard2019judgment,martelaro2020could,shen2021value} to support speculation and conversation about the potential impacts and harms of AI.
Such activities can inform practitioners about common harms (including but not limited to fairness-related harms, such as discrimination, and privacy-related harms, such as undue surveillance), but they may be too broad to help them imagine concrete consequences they have not seen before.
While we do not doubt the value of these tools, we believe their effectiveness may require participants to have significant background knowledge of potential harms going in, establishing the need for practitioners to develop a more extensive awareness of specific historical incidents of harm. Such awareness would help teams prevent repeat mistakes and offer them a more solid foundation when speculating about the downstream harms of their creations.

Taking an \emph{ethics-first} approach, we envision a future in which such awareness is instilled in AI students from an early stage through carefully designed and validated educational tools---long \emph{before} they start their careers as AI practitioners and creators. 
Prior work has likewise emphasized the \emph{awareness of harmful consequences} as one of the key learning objectives of data science ethics~\citep{saltz2018key,fiesler2020we}.
However, as \citet{fiesler2020we} observe, \qt{Learning outcomes in [tech ethics] classes focused far more on conceptual skills than on specific knowledge;} in particular, \qt{many topics within tech ethics are high level and conceptual when it comes to the impact of technology on society.} 
While we are motivated to eventually help students overcome a failure of imagination with potential harms, we believe that the first step in addressing this is to raise student awareness of past AI harms.
In this work, we set out to propose and test an educational tool to expand students' awareness of AI harms using tangible historical examples.

\xhdr{The AI Incident Database}
To help AI system developers learn from prior failures, several groups have begun compiling lists of AI issues and incidents~\cite{charles_pownall_ai_2019,david_dao_awful_2021}. 
One of the most mature lists in this space is the AI Incident Database (AIID)\footnote{https://incidentdatabase.ai}, which provides a listing of over 2000 incident reports for over 400 incidents where AI has caused or nearly caused harm~\citep{mcgregor2021preventing}.
The AI Incident Database takes inspiration from the NTSB Aviation Accident Database \cite{ntsb_ntsb_nodate} and provides a rich set of metadata about incidents that can be useful for researchers, product managers, and engineers searching for and analyzing potential harms based on prior incidents.
Across the database, major categories of AI harms are reported for autonomous robots, language/vision models, autonomous driving, recommender systems, identification systems, AI supervision, and healthcare systems \citep{wei2022ai}.\footnote{The reader may find it interesting to know that we relied on AIID to find and summarize recent harmful incidents involving computer vision, large language models, and autonomous vehicles, referenced at the beginning of this introduction.}
In addition to being a relatively comprehensive listing of recent AI incidents, AIID provides a fairly usable interface, including a set of filters and search capabilities, making it a potentially effective tool for classroom instruction. To our knowledge, there has yet to be any research assessing the use of the database in classroom environments and its efficacy in raising students' awareness of harmful AI incidents.

In this work, we propose an educational module built around AIID with the objective of raising awareness of AI harms as part of the AI ethics curriculum. Our proposed module consists of a pre-activity questionnaire, two in-class components, and a post-activity questionnaire, organized as follows:
\begin{enumerate}
    \item \textbf{Pre-activity questionnaire} to assess students' initial familiarity with AI harms and related AI ethics topics
    \item \textbf{In-class activity} to explore incident records in AIID
    \item \textbf{Post-activity questionnaire} to assess the impact of the activity on students' awareness of AI harms
    \item \textbf{In-class discussion} to reflect on the experience using AIID and report the findings of the questionnaires
\end{enumerate}
We present findings obtained through a classroom experiment conducted in an R1 institution as part of a course on the societal and ethical considerations around using AI and ML in society. 
We observe that before interacting with AIID in class, students mainly learned about AI news through social media. However, afterward, they found AIID to be a valuable resource for periodically updating their knowledge of AI incidents. Interacting with the database helped them better understand the \emph{magnitude and severity} of AI harms in the real world and instilled in them a sense of urgency around (a) designing \emph{functional and safe} AI systems and (b) strengthening \emph{governance and accountability} mechanisms to prevent and litigate realized harms.

This shift in mindset is critical, especially when viewed in connection with recent reports contrasting the values and priorities of various AI stakeholder groups: \citet{jakesch2022different} have found that AI practitioners prioritize fairness over safety even though estimates of the general US populace hold the opposite view. \citet{laufer2022four} report growing calls from within the AI ethics community to prioritize accountability over fairness. Further, \citet{raji2022fallacy} liken the proliferation of poorly functioning AI (of which there are many) to that of faulty, unsafe products. In food industries, financial services, and consumer electronics, people have demanded accountability and regulation to assure safety.
In light of these reports, we believe it is important for ML experts to have value priorities in line with those of other stakeholders and to be aware of the harms of the technology on the ground.
We also view the shift in students' mindsets to thinking critically about the functionality of AI systems and mechanisms of accountability as improved alignment with other stakeholders with the potential to amplify demands for governance mechanisms (such as consumer protection rights) that have made other industries much safer and more reliable.
Students are future technologists and practitioners, so their awareness of potential harms is a first and major step in moving the industry toward safer implementations and systems to protect people.
Thus, as part of this work, we look to see how students' value priorities might change after exposure to the AIID.

\xhdr{Limitations and future directions} This work is an initial attempt toward scientifically designing a set of educational tools to achieve the learning objectives of AI ethics curriculum development. While we believe awareness must be an early and indispensable goal of any such effort, as we outline in Section~\ref{sec:analysis}, other objectives, such as \emph{effective engagement with other stakeholders}, \emph{critical thinking}, \emph{principled ethical reasoning} are equally crucial, but not the focus of our class activity. The findings reported here result from a small-scale, time-restricted educational activity conducted in a single institution. While these limitations are common in prior classroom studies (see, e.g., \citep{kasinidou2021agree,Pierson_2018}), we note that for the results of such small-scale studies to be generalizable, they must be repeated with larger sample sizes and in multiple environments (including in other educational institutions and ultimately, in practical data science workflows).
We also emphasize that our study evaluates a specific educational activity with AIID and not the database itself, meaning results could differ if an alternative activity with the same database was employed instead.
Despite its limitations, we hope this work can inform future usages of AI incident databases in educational and practical settings and lead to concrete improvements to the accessibility and reach of existing tools. Section~\ref{sec:discussion} contains a list of actionable recommendations for the AIID team and the broader community to enhance the usage and impact of AI incident databases. 

\subsection{Related Work}\label{sec:related}
Integrating ethics into computer science curricula has a history spanning several decades (see, e.g.,~\citep{nielsen1972social,martin1996implementing}). Notable recent developments include Harvard's Embedded EthiCS~\citep{grosz2019embedded} and Stanford's interdisciplinary course on Computer Ethics~\citep{reich2020teaching}. For an in-depth description of current trends in computing ethics education, we refer the reader to Feisler, Garret, and Beard's \cite{fiesler2020we} analysis of tech ethics in CS course syllabi. 

While AI ethics modules and courses have become popular in recent years, we still need a coherent set of scientifically designed and evaluated educational tools, activities, and modules developed to achieve well-specified learning objectives. For example, \citet{bates2020integrating} shed light on the challenges of integrating issues of fairness, accountability, transparency, and ethics (FATE issues) into data science curricula, as witnessed by the teaching staff of an MSc-level course, and concludes with raising several fundamental questions: ``\textit{Without creating too many restrictions on teaching practice and innovation – what should a well-designed [Ethics of] Data Science curricula address? What should curriculum designers be thinking about, and what should students understand and be able to do at the end of their studies? How might educators go about teaching key and emerging topics?}'' [pg. 10].
Other research in the AI education area reveals that students' backgrounds influence their views on AI ethics, suggesting that a ``one-size-fits-all'' approach to instruction may not suffice. \citet{kasinidou2021agree} find that CS students' levels of education affect how they interpret the ethics of algorithmic decision-making outcomes. \citet{Pierson_2018} describes how students' demographics such as gender can also play a role in AI ethics judgment. These findings motivate the need for effective ethics curricula as part of AI education.


Prior work has offered and evaluated educational tools and activities for teaching AI ethics~\cite{skirpan2018ethics,saltz2019integrating,Kasinidou_Kleanthous_Orphanou_Otterbacher_2021,Pierson_2018}, some designed specifically for learning objectives such as \emph{perspective taking}~\citep{shen2021value,brown2022shortest}, \emph{developing care}~\citep{shapiro2020re}, \emph{reflexivity}~\citep{cambo2022model}, and \emph{contextualizing ethical issues}~\citep{bullock2021computing}. Other contributions focus on specific values such as \emph{fairness}~\citep{mashhadi2022case} and \emph{reproduciblity}~\citep{lucic2022reproducibility}. To our knowledge, no prior work has proposed or evaluated teaching tools for the learning objective of \emph{AI harm awareness}.

Among broader engineering ethics education, analyzing well-known engineering failure case studies (e.g., Challenger disaster, Chernobyl explosion, and Exxon Valdez oil spill) are common tools to engage students in ethical thinking \cite{colby2008ethics}.
Such case studies help students realize the potentially grave harms of engineering mistakes or poor design decisions.
AI ethics education is starting to incorporate similar case studies. 
For example, in a survey of US AI Ethics courses, \citet{Garrett2020More} found that many classes included reading news articles about AI incidents, suggesting the activity as a way to help students connect with the real-world impacts of AI.
Further, exposure to past harms can help build up students' mental library of possible damages that may occur.
However, among engineering ethics education in the US, there is still little empirical work studying the impacts of such case study methods and best practices for implementing classroom activities to support ethical learning goals \citep{hess2018systematic}.
In this study, we look to see how engaging students using the AI Incident Database can raise their awareness of possible AI harms and enrich their overall thinking about AI ethics.

\section{Research Design and Methodology}\label{sec:methodology}
The research activity had an in-class component (80 minutes long) and two out-of-class components (estimated to take ~30 minutes to complete). All students enrolled in the course were required to participate in the activity to receive points for participation and homework assignments, but they could opt out of their data being used for research purposes. Opting out did not impact students' grades or course outcomes. (Appendix~\ref{app:consent}
describes the organization of the consent and data collection processes to ensure students didn't feel coerced to take part in the study.)

\xhdr{Data collection}
Data collection proceeded in four steps: 
\begin{enumerate}
    \item Before the first course lecture in which AIID  was introduced, the study participants completed a brief questionnaire assessing their initial familiarity with the relevant concepts and eliciting some background information about them. (This step is outlined in Section~\ref{sec:pre-class}.)
    \item During the classroom activity, the instructor introduced students to AIID and prompted them to work with the tool (a) individually and (b) in teams and submit an incident report to the Database. (See Section~\ref{sec:class_activity} for further details.)
    \item After working with AIID and submitting a report, the instructor led a class-wide discussion encouraging students to reflect on their experience with and potential limitations of the tool. (See Section~\ref{sec:AIID-feedback}.)
    \item Within three days after the in-class activity, participants were asked to complete a post-class questionnaire designed to assess the impact of the activity on their perceptions.
\end{enumerate}

\xhdr{Participation consent and benefits} The study did not ask students to do anything beyond the normal activities and assignments that were integral to the course. They were informed that if they decided to participate, an anonymous version of their data would be shared with the research team. Students could opt out of their data being used for research purposes. Participants did not receive any compensation (either monetary or in terms of course grade), and opting out did not adversely impact course outcomes in any way. As an indirect benefit to students, the scientific knowledge produced by the research was shared with the class immediately after the data collection and analysis were concluded.

\xhdr{Population and sample} The population of interest was students with a background in Machine Learning interested in learning about the societal aspects of the technology. Our study participants were students enrolled in a class designed to address these topics. Since the course was not a degree requirement for any major and having completed an ML course was a prerequisite, we expect the sample of students who registered for the course to be roughly representative of the population of interest. Participation in the study was limited to students aged 18 and older. 
Our participants consisted of mostly STEM-major university students, therefore, as one may expect, the sample was biased toward
young and liberal students. This is consistent with our population of interest.
We did not target any minority demographics specifically, but participants could voluntarily provide us with their basic demographic information.

\vspace{-2mm}
\subsection{Pre-class Questionnaire}\label{sec:pre-class}
The pre-class questionnaire (Appendix~\ref{app:pre-class-questionnaire}) aims to understand students' background, knowledge, and goals as they relate to Machine Learning (ML), Artificial Intelligence (AI), and the topics of AI use in society covered in the course. It additionally asks students to provide information about their educational background and demographics. The purpose of these questions was to detect any significant variations in responses across the corresponding dimensions. 

The questionnaire (which took 10-15 minutes to complete based on our pilot) consisted of the following three sections: 
\begin{itemize}
    \item Understanding student's \textbf{educational background} and motivation for taking the course;
    \item Assessing students' initial familiarity with the use of \textbf{ML in society} along with their beliefs regarding which values were most important to prioritize, as summarized in Table \ref{tab:value-descriptions} and \citep{jakesch2022different};
    \item Understanding the sources through which they learn about \textbf{AI news} and advances.
\end{itemize}

\vspace{-2mm}
\subsection{In-class Activity}\label{sec:class_activity}
The in-class activity was organized as follows (details can be found in Appendix~\ref{app:in-class-activity}):

\noindent\textbf{(1) Session overview:} The instructor provided an overview of the session’s goals and the importance of a shared database containing prior AI incidents. The instructor then introduced the AI Incident Database, focusing on the ``tabular view.''

\noindent\noindent\textbf{(2) Individual activity:} Individual students were then assigned a number 1--40 and were asked to explore the 10 incidents on their corresponding page of the database's table view (to prevent students from only observing most recent issues). Students then chose one incident that caught their attention (e.g., because it’s news to you, it’s surprising, the magnitude of impact could be significant) and responded to the following questions about the incident report (student answers can be found in Appendix~\ref{app:in-class-findings}):
    \begin{itemize}
        \item What was the \underline{source} of the story (e.g., which website/author published it)?
        \item In what \underline{application domain} did the incident occur?
        \item What was the \underline{nature of the incident} (i.e., the concern that was raised)?
        \item Who was (partially) \underline{responsible} for the incident (e.g., because they developed, deployed, or used the AI system)?
        \item Who was (potentially) \underline{harmed}?
        \item How was the incident ultimately \underline{addressed} (Was there a penalty? Was the tool discontinued?)
    \end{itemize}

\noindent\textbf{(3) Team activity:} Students then formed teams of four and were asked to look for a recent (i.e., 2019–2023) AI/ML incident story that was not documented in the database. Students were directed to the ``Spatial View'' of the database if they needed inspiration for areas to look for new incidents. The team then submitted a report on an undocumented story to the database.

\noindent\textbf{(4) Discussion of limitations:} The instructor moderated a class-wide discussion on the challenges and limitations of the database, using the following prompts:
    \begin{itemize}
        \item Did you find AI incident news sources that were typically not covered by the database? Why?
        \item Did you find the database up-to-date or outdated? Why do you believe that is the case, and what can be done about it?
    \end{itemize}

\vspace{-2mm}
\subsection{Post-Activity Questionnaire}\label{sec:post-activity}
The post-activity questionnaire (Appendix~\ref{app:post-activity-questionnaire}) aims to assess the efficacy of the class activity by eliciting students' feedback about the tool and asking them to revisit the questions they encountered in the pre-class questionnaire. In particular, it asks students whether working with AIID impacted their response to the questions about:
\begin{itemize}
    \item their \textbf{motivation} for taking this course and the topics they want to learn about,
    \item the \textbf{applications} of ML in society that they are excited/concerned about, 
    \item the \textbf{values} they believe should be prioritized, as summarized in Table \ref{tab:value-descriptions} and \citep{jakesch2022different},
    \item and the \textbf{role of ML experts} in promoting those values. 
\end{itemize}
\section{Analysis and Findings}\label{sec:analysis}
We ran the study outlined in Section~\ref{sec:methodology} with the approval of our Institutional Review Board (IRB).\footnote{The corresponding CMU IRB number is STUDY2022\_00000447.} 
The in-class activities were performed during the first lecture of a graduate-level course on societal and ethical implications of AI/ML offered as an elective by the Computer Science unit of an R1 institution.\footnote{According to the Carnegie Classification of Institutions of Higher Education, R1 is defined as ``doctoral universities'' with ``very high research activity''.} 
We elicited students' feedback about AIID and measured their learning outcomes before and after introducing the tool to assess its efficacy in improving students' awareness of AI incidents, limitations, and areas for improvement. We did not use a control group, instead opting for a pre-vs-post, within-groups study and follow-up analysis.
In total, 36 students participated in at least one questionnaire, and of those, 30 participated in both. We emphasize again that because our sample is small and from a single institution, our results should not be presumed to generalize to other student populations. Rather, they serve as suggestive evidence to inform future studies of similar educational activities using AI incident databases. That said, we note that small sample size is a common issue in prior work on AI ethics classroom activities (see, e.g., \citep{kasinidou2021agree,Kasinidou_Kleanthous_Orphanou_Otterbacher_2021,Pierson_2018}).

\xhdr{Participants' characteristics}
Figure \ref{fig:participant-distribution} in Appendix \ref{app:demographics} contains the full demographic breakdown of our participants.
With regard to certain characteristics, our participant group is relatively diverse. For instance, our group exhibits a 50-50 split across gender lines, and while ``Liberal'' was a majority answer for political views and a majority of students didn't belong to a marginalized group, at least 25\% of students answered differently in each case. With regard to other characteristics, our participant group was highly homogeneous. Most students came from a background in Science, Technology, Engineering, and Mathematics (STEM), and all were less than 40 years old. We discuss how these observations affect the generalizability of our results in Section~\ref{sec:discussion}. 

\xhdr{Thematic analysis} Many of our survey questions required participants to submit free-form, open-ended responses. To extract the notable themes from responses, we used the following process: For each question, we read through all submitted responses, created initial codes, and grouped similar codes together as potential themes. We reread all responses to revise themes as necessary. Given the small number of participants, we did not end up making any significant changes to our themes at this stage. We took all extracted themes into account to construct the narrative presented here.

\xhdr{Reflections on positionality}
As is always the case with qualitative research, our background and positions as researchers and instructors have undoubtedly influenced both the data we collected and our interpretation of it. Our team comprises three researchers: two in senior and one in junior roles. Two members of the research team also served as teaching staff for the course in which the study was conducted. None of the students in the course opted out of the research study. Even though we had ensured that the course instructors wouldn't know the identity of the students opting out, students may have felt obliged to support the instructor's research efforts. The same context may have impacted their responses.
Collectively, our team represents a variety of gender and racial identities, as well as cultural backgrounds and national origins. We are currently affiliated with academic institutions in the US. Our past experiences and perceptions regarding the lack of clearly-defined learning objectives and scientifically-evaluated educational tools in the AI ethics space motivated us to undertake the current study. Lastly, we analyzed the qualitative data to deliberately pull out comments that informed our research questions. The narrative presented here does not highlight many instances of generic or otherwise non-informative responses.\footnote{With permission from our IRB, we plan to publish an anonymous version of our raw data so that other researchers can form their own interpretations.}

\subsection{Educational Motivations, Goals, and Interests}
As demonstrated in Figure~\ref{fig:participant-distribution} in Appendix~\ref{app:demographics}, the majority of participants came from STEM backgrounds, were enrolled in a graduate-level science or engineering major, and characterized their familiarity with AI/ML as intermediate to advanced. However, most participants (73\%) had no prior AI ethics training. Only 27\% of participants had some form of AI ethics training in the past. 8\% had taken general AI ethics courses; another non-overlapping 8\% had taken specialized AI ethics courses, such as ones focusing on privacy, environmental sciences, and natural language processing; 11\% had briefly learned about topics such as fairness as part of their regular CS/AI courses.

\xhdr{Motivation for taking the course}
 Collectively, students stated they took the course out of a sense of \texttt{personal responsibility}, and to be able to anticipate ethical issues that may come up in their \texttt{future careers} as they apply ML to real-world problems. For example, one participant said \qt{I think learning about ethical usage of ML is an important responsibility for people planning to pursue ML. [S23]} Some expressed a desire to bridge the \texttt{educational gap} they perceived between their technical skills and their ethical prowess, with one student stating that \qt{ethics are important to me, but the application of ethics to ML and similar things is not obvious. [S28]} Some took the course due to their interests in special \texttt{research topics}, such as in fairness, privacy, and explainability, or specific applications of ML in domains such as medicine, policy development, and autonomous driving. Finally, several students expressed a desire to develop \texttt{connections with like-minded students}.
 
\xhdr{Learning goals}
Students mentioned several learning goals related to building more familiarity with AI ethics topics. Multiple students expressed a desire for a \texttt{systematic, principled conceptual framework} to identify, evaluate, and address ethical concerns around AI. For example, one participant said \qt{I hope to gain some more organization and structure when thinking about how ML affects society and vice versa. [S2]} Another said \qt{I hope to gain a principled understanding of current theory about ethical usage of ML and tools or benchmarks applied to evaluate such usage. [S23]}
A different frequently mentioned goal was raising \texttt{awareness} of the type of harms and issues they should foresee when deploying AI in high-stakes domains. For example, one student said \qt{I want to set a radar for myself so that I'm aware of the potential ethical considerations when I design/train models in the future. [S13]} Another expressed interest in learning about \qt{considerations I may not have thought of independently. [S6]}
Some students were interested in learning about \texttt{best practices}, including learning about \texttt{concrete tools and approaches} to address ethical considerations, such as those designed to detect and mitigate bias in ML systems or explain their inner workings. For instance, one student said they were interested in learning \qt{A more in-depth understanding of what value alignment looks like in practice [...] and how we can integrate these conversations about ML ethics into large-scale models deployed in industry today. [S19]}
Strengthening \texttt{critical thinking} skills to address tradeoffs between competing values was another common theme. For example, one student said: \qt{I want to be able to produce logical arguments in the context of machine learning and be able to think about such topics critically. [S16]} Another mentioned \qt{I hope to be able to get a hands-on experience in doing a critical study on popular AI algorithms [...]. [S12]}
Students also reported wanting to practice \texttt{constructive dialogue} to better understand and respond to people with other perspectives from their own. 
While the main focus of the current work is on raising awareness and expanding students' mental library of potential issues, we believe that the additional themes reported here are equally important learning objectives for AI ethics curricula.

\xhdr{Topics, tools, and applications of interest}
Regarding specific topics, tools, or applications of interest, at least six students expressed a desire to learn more about the ethical challenges around \texttt{generative models} such as ChatGPT and deepfakes. Many students reported interest in learning about \texttt{fairness and de-biasing toolkits}. At least two specifically mentioned IBM's AI Fairness 360 package \citep{bellamy2019ai}. \texttt{Explainability} and transparency tools were mentioned several times but without details about the specifics. Only one student mentioned LIME \citep{ribeiro2016should} and SHAP \citep{NIPS2017_7062} as concrete methods they would like to learn about. Among applications mentioned, only \texttt{medicine/healthcare} was mentioned by more than one student. Notably, none of the students mentioned \textit{safety} as a topic of interest in response to the question about topics, tools, and applications.

\xhdr{Exciting vs. concerning use cases of AI in society} In response to specific applications of AI in society that they are \emph{excited} about, students mentioned AI in \texttt{medicine} (e.g., for personalized care, helping professionals with diagnosis and prognosis of diseases, managing public health crisis such as the COVID pandemic, and drug discovery), \texttt{environmental protection} (e.g., climate modeling and monitoring deforestation), \texttt{transportation} (e.g., traffic routing), \texttt{finance} (e.g., to perform autonomous stock trades), \texttt{language translation}, and evidence-based \texttt{policy making}. 
In response to AI use cases they are particularly \emph{concerned} about, students mentioned \texttt{generative models} (e.g., copyright concerns around generative arts and deepfakes, the negative impact of ChatGPT on education), \texttt{surveillance and profiling} (e.g., in targeted advertising, facial recognition in public spaces, and prescription drug monitoring), \texttt{military uses} (e.g., autonomous weapons), \texttt{criminal justice}, and \texttt{employment} (e.g., surveillance, replacing human jobs). Two use cases, (i.e., autonomous driving and public safety) garnered mixed responses, with some students considering them problematic, while others saw their potential for positive impact.

Interestingly, our inspection of how students learned about the above use cases revealed that the majority of respondents (i.e., 23) learned about positive/exciting use cases through educational material (e.g., courses and research opportunities), but came across negative/concerning applications mainly through social media and personal encounter (26 instances), and less frequently through educational institutions (only 12 mentioned sources that could be plausibly attributed to this category). This observation has clear implications for AI educators and reinforces the cautionary message that we must be deliberate in painting a realistic picture of the technology for students~\citep{smith2022real}.

\xhdr{Role of ML experts and practitioners} In response to the question about the role of ML experts and engineers in promoting responsible AI values, students frequently brought up the need for a shift in \texttt{industry culture} to foster awareness, reflexivity, diversity, and continual training. For example, one student suggested that ML practitioners must \qt{apply [ethical thinking] to their work, bring up and discuss it in group/team meetings, ask company/institution to provide training on it [...]. [S39]} Another said \qt{I believe that the first step for MLEs to be able to promote fairness in their models is to ensure that they receive the proper education/training/awareness of these issues especially when the model will be deployed in sensitive fields such as healthcare and education. [S19]} 
Related to \texttt{diversity in ML teams}, one student highlighted the imperative to \qt{[...] first recogniz[e] their internal biases and then creat[e] teams where the people developing these programs themselves are diverse in their backgrounds. I think it can be really difficult to catch your biases when developing a programs because you may be preconditioned to believe your data and the results you're expecting should look a different way, but by having engineers of different backgrounds you have people who may catch your biases and you can catch theirs. [S29]} 
Consistent with their aforementioned desire for more structure and systematic frameworks in place, multiple students emphasized the need for strengthening \texttt{best practices, standards, and tools}: \qt{Experts and Engineers could create a set of safety guidelines that every AI or ML machine must meet for it to be used or deployed. [S24]} Investments in improving \texttt{data quality} and conducting more \texttt{extensive testing} were among the concrete suggestions that were brought up by participants: \qt{More time need[s] to be devoted to testing the AI system under various circumstances to ensure that it would be safe to operate. [S3]} One participant noted the difficulty of such testing by providing an example: \qt{In self driving cars for example, engineers [have to] use data on all types of terrain to train the car so that it is able to safely navigate these areas. [S36]} Several students mentioned additional \texttt{technical remedies}, including \qt{designing new algorithms/optimization objectives [S32]}; \qt{[understanding] which part of the model is responsible for a particular 'unfair' decision? Do the weights need to be normalized with respect to a variable? [S12]}; and \qt{incorporat[ing] physics models into the AI algorithms, such that the model can be more interpretable. [S27]}

A small number of students (fewer than 3) mentioned stronger \texttt{engagement with non-technical stakeholders} (e.g., experts in humanities and social sciences, the public, impacted community members, and policy makers) as a necessary step toward promoting values such as fairness: \qt{I think that we as a community can do a better job engaging with ongoing social movements and important research on technology and society that exists in our field.  The most influential works that I've read that have shaped my opinions on technology \& society have some from sociology, anthropology, history, etc.  We can contribute by listening to, supporting, and collaborating with these people [S9].} The same participant emphasized the need to go beyond personal responsibility, and develop systems of \texttt{accountability and governance}: \qt{even if every ML expert is individually trying their best to do what's ``ethical'', the US needs to regulate corporations that deploy AI.  We could perhaps help by raising public awareness of our work.  I feel like we as a community under-emphasize scientific communication/public engagement about our work (because it's not incentivized in the academy) [S9].} 

\vspace{-2mm}
\subsection{The Effect of AIID on Awareness}
\xhdr{Changes in learning motivation}
66.7\% of students reported that working with AIID increased their motivation for taking the course. For example, one student mentioned \qt{I did not realize there were so many incidents of AI and ML going wrong in industries where I did not expect and it makes me want to learn more. [S29]}
The remaining 33.3\% reported it didn't impact their level of motivation, because they were highly motivated to begin with and already aware of the examples they saw in the database. For example, one student said \qt{I was already aware of many of the types of incidents covered in the database. [S23]} No one reported a decrease in motivation as the result of the class activity.

\begin{figure}
    \centering
        \begin{subfigure}{0.29\textwidth}
        \includegraphics[width=\textwidth]{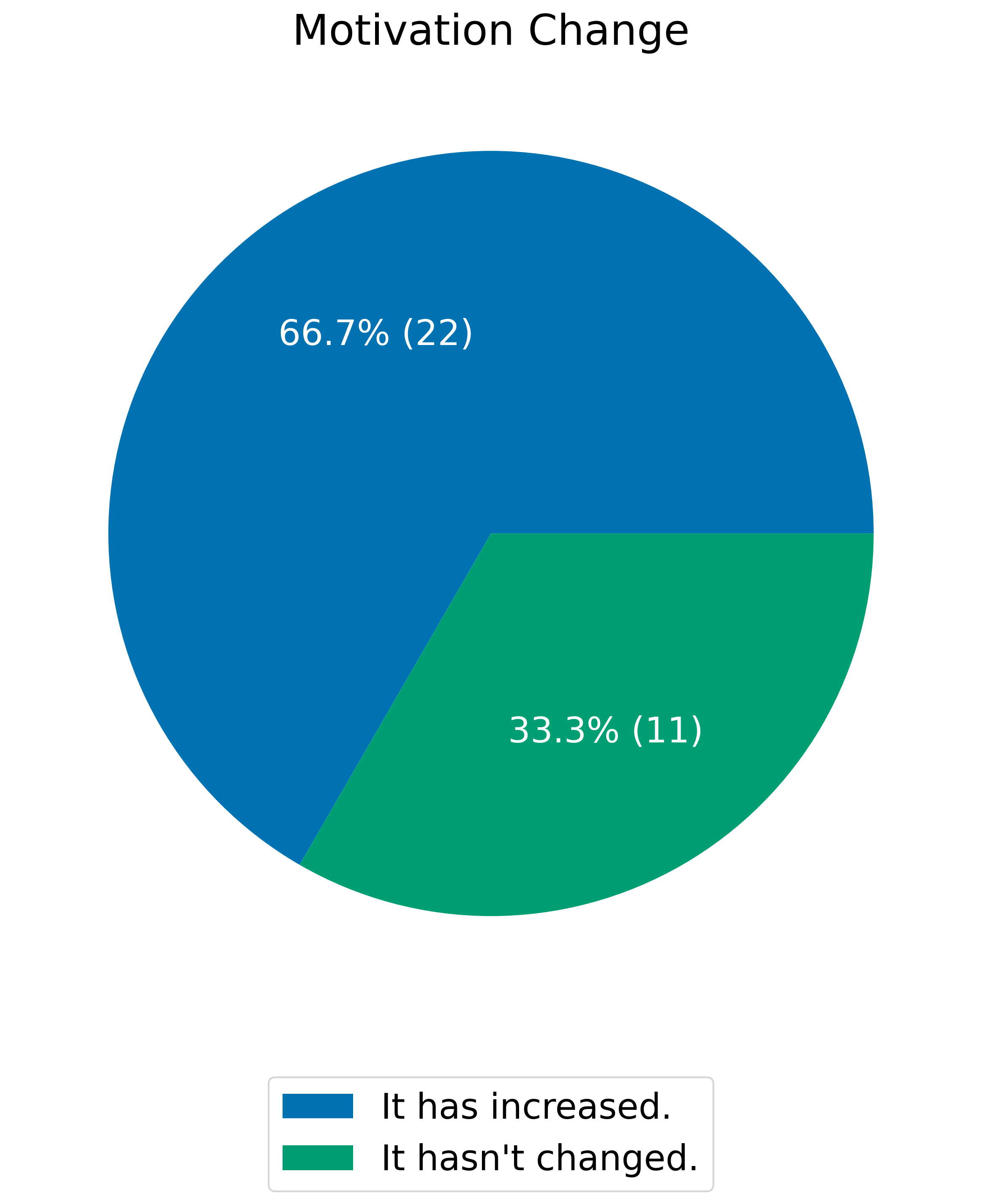}
    \end{subfigure}
        \hspace{0.14\textwidth}
    \begin{subfigure}{0.5\textwidth}
        \includegraphics[width=\textwidth]{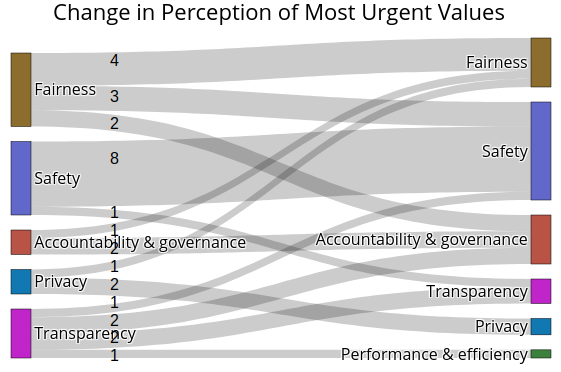}
    \end{subfigure}
    \hspace{0.1\textwidth}
    \vspace{-4mm}
    \caption{Sankey diagram and pie chart illustrating students' changes in beliefs as a result of interacting with the database based on responses from the 30 students who completed both surveys. Overall, the intervention primarily shifted students' priorities towards issues of ``Safety'' and ``Accountability \& governance'' at the cost of focus on ``Fairness'', ``Privacy'', and ``Transparency''. Additionally, 66.7\% of students reported an increase in motivation after working with AIID. We conjecture that the saturation of the database with reports such as autonomous vehicle accidents and data misuse could have caused these shifts.
    }
    \label{fig:changes}
    \vspace{-3mm}
\end{figure}

\xhdr{Changes in priority of values}
Both before and after the AIID activity, we asked students via our surveys about which value related to real-world deployment of AI was the most urgent one to address in their views. Using responses from the 30 students who completed both surveys, we plotted changes in perceptions of these values as a result of the survey using a Sankey diagram and pie chart, shown in Figure \ref{fig:changes}. Before the activity, Fairness and Safety were students' most urgent values to prioritize to promote responsible use of the technology. However, the activity appears to have caused a shift towards Safety and Accountability \& Governance and away from Fairness, Privacy, and Transparency. 
Further analysis of students based on whether they had prior AI ethics training reveals that of the 10 students with prior training, half of them chose Accountability \& Governance after the activity. In contrast, only one student out of the 20 students lacking such training chose Accountability \& Governance afterwards. Another breakdown across political lines shows that 6 of the 22 self-identified left-leaning students chose Accountability \& Governance after expoloration of AIID while \emph{none} of the remaining 8 self-identified non-left-leaning students chose that RAI value after the exploration.  
While we caution against attributing statistical significance to our results or making
causal claims due to our sample size (see Section~\ref{sec:discussion} for more details), we perform statistical tests of overall marginal homogeneity and report our results in Appendix~\ref{app:tests} alongside an interpretation of our p-values. We conjecture that these shifts in priorities are due to the prevalence of incidents related to poor AI functionality and safety (e.g., autonomous vehicle accidents), and the observation that in most cases there were no serious ramifications for wrongdoers. This conjecture is supported by the observation that most of the issues found interesting by students pertained to Safety defined in \citep{jakesch2022different} (as identified in Appendix~\ref{app:in-class-findings}), but we leave formal verification of this conjecture to future work.

Qualitative coding of open-ended survey responses and feedback offered during the class discussion seem to support the quantitative conclusions here. Students wrote that AIID was helpful in learning about \texttt{new incidents}, some of which were \texttt{unexpected/shocking}. They also wrote that the incidents reported in AIID stressed the importance of \texttt{safety} when AI is deployed and gave them better senses of \texttt{magnitude} and \texttt{severity} of AI issues. As part of the follow-up discussion, students brainstormed various approaches towards accountability. While students agreed that \qt{action should be taken to make sure an incident does not repeat}, they offered different solutions as well as reasons why taking action is challenging. Some mentioned \texttt{risk-management} frameworks such as \qt{datasheets for datasets} \citep{gebru2021datasheets}. Others considered more \texttt{proactive} and \texttt{regulatory} strategies such as via \texttt{human-value alignment} or \texttt{human-in-the-loop feedback}. Still others pointed out that a \texttt{lack of explainability} and \texttt{speed of development} make these problems difficult, but \qt{thinking about what we want to happen} could be a step forward.

\begin{figure}
    \centering
     \begin{subfigure}{0.29\textwidth}
         \centering
         \includegraphics[width=\textwidth,valign=t]{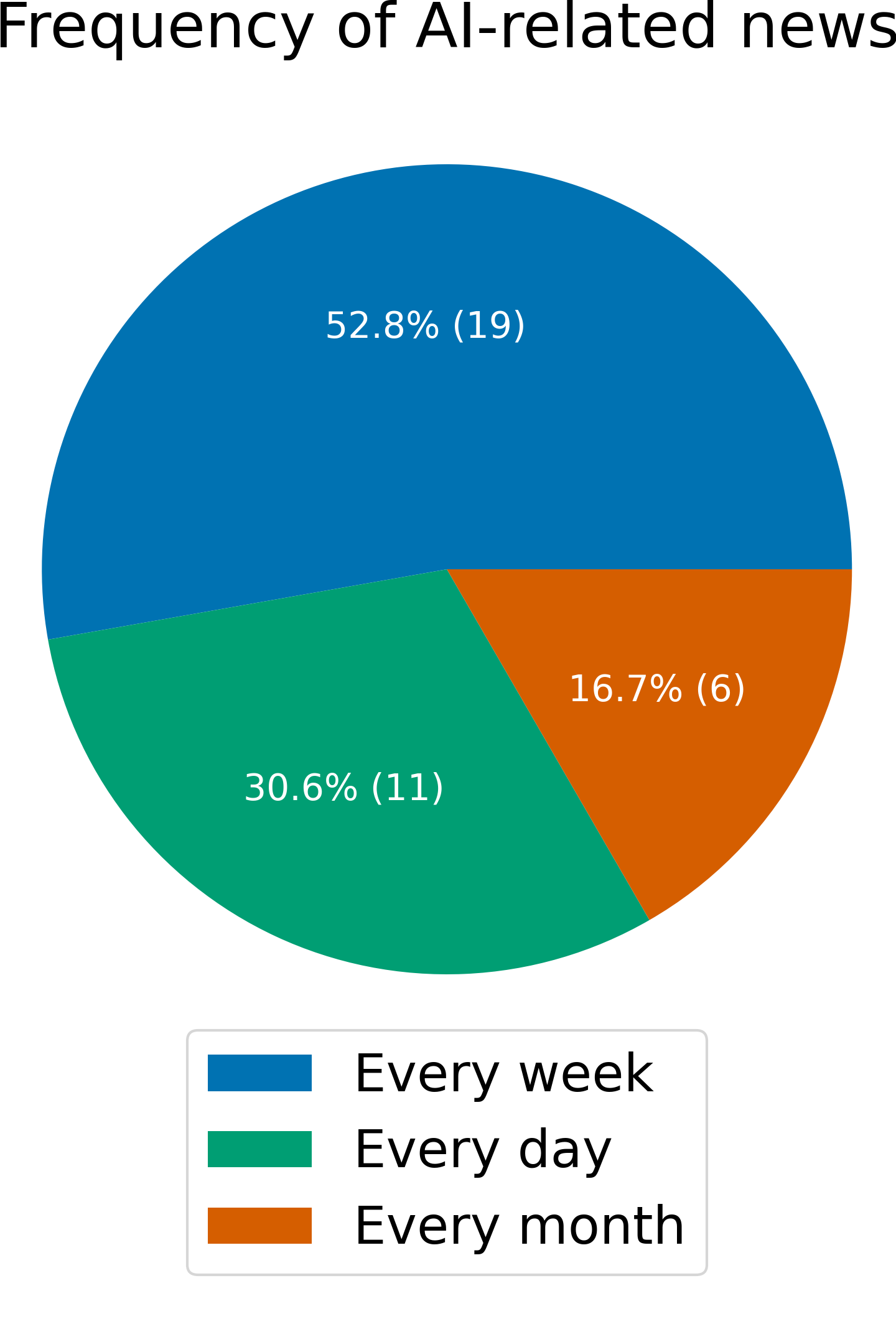}
     \end{subfigure}
     \hfill
     \begin{subfigure}{0.27\textwidth}
         \centering
         \includegraphics[width=\textwidth,valign=t]{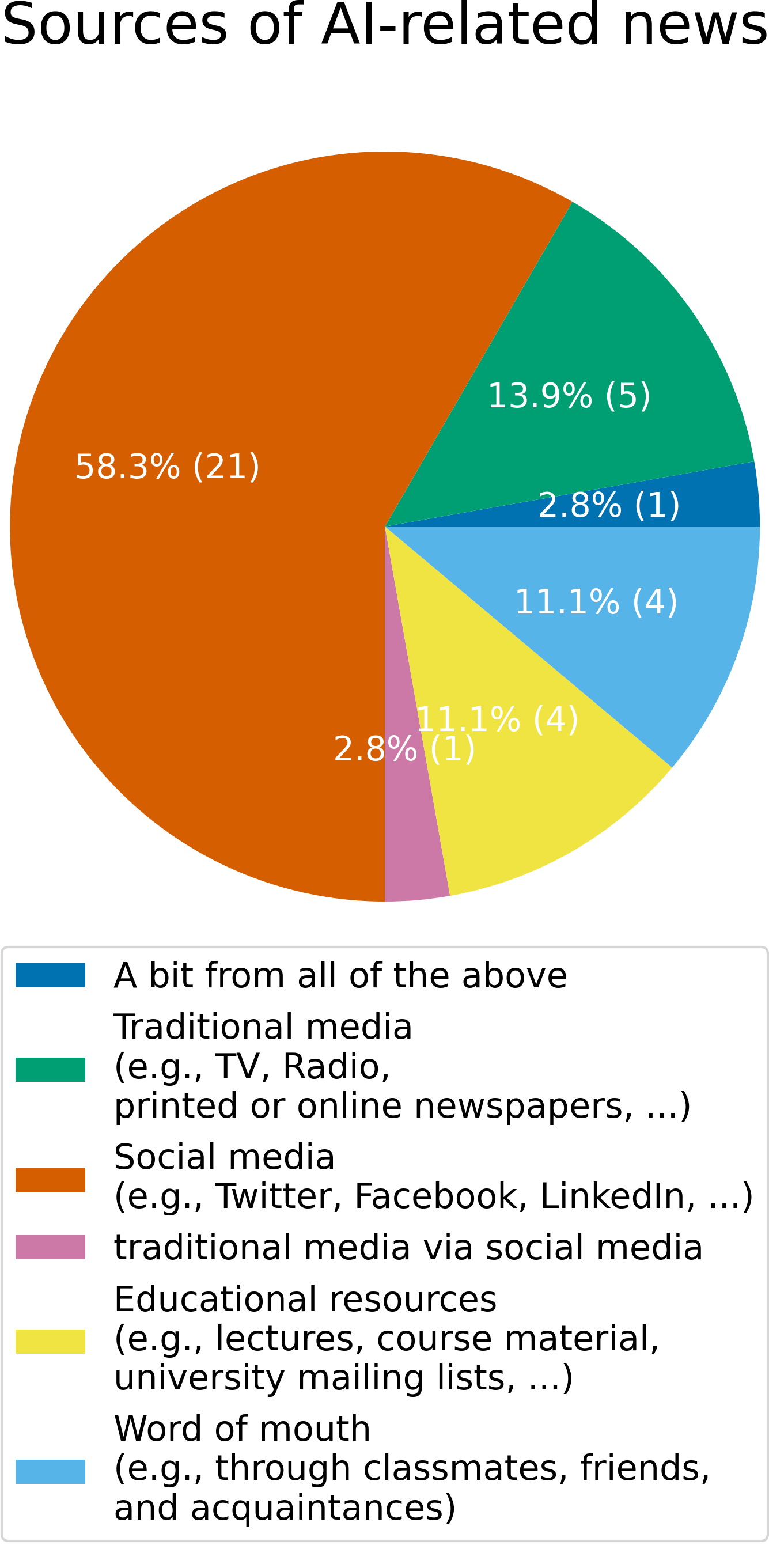}
     \end{subfigure}
     \hfill
     \begin{subfigure}{0.32\textwidth}
         \centering
         \includegraphics[width=\textwidth,valign=t]{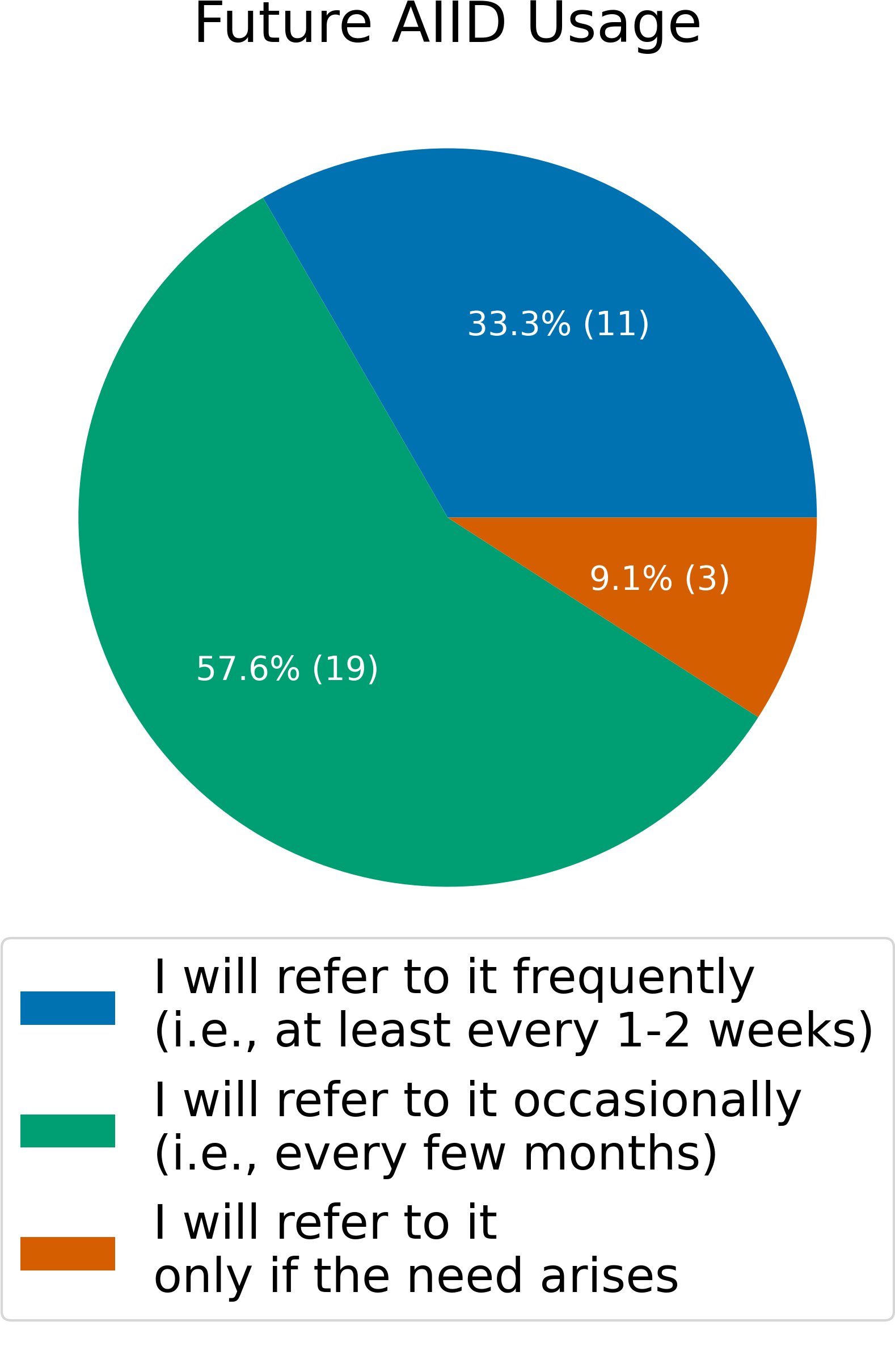}
     \end{subfigure}
     \vspace{-4mm}
    \caption{AI-related news report habits of the students who participated in our study. We find that most students tend to hear about AI-related news every week. Additionally, they typically learn of developments through a mixture of social media and traditional media, though the former is most common. After our study, most students reported that they would refer to the AIID occasionally or frequently, suggesting that the resource is useful but may warrant some improvements. Recommendations are covered in our Discussion section.}
     \label{fig:source-distribution}
\end{figure}

\xhdr{Changes in sources of AI-related news}
We asked students about how they gathered AI-related news before encountering AIID and whether they would use AIID in the future. Pie charts illustrating the distribution of these reponses are shown in Figure \ref{fig:source-distribution}. Most students learned of AI-related news every day or every week and typically through a mixture of social media and traditional media. Upon learning about AIID through the class activity, most students said they would use it frequently or occasionally, suggesting that found AIID a useful resource overall.

\xhdr{Changes in topics and tools of interest} When asked about whether there were any additional tools and topics of interest they would want to explore as a result of interacting with AIID, seven students reported that there were not, one student did not respond, and another confessed they were \qt{still unsure. [S5]} All other students responded affirmatively. Four of them reasoned about policy and accountability: \qt{I think I am very much interested in policy and how to solve the the problem because there a lot of issues but none of them are really resolved. [S37]} Two students mentioned wanting to learn more about testing and evaluation of AI models, such as \qt{ways to identify incidents or test AI algos [S32]} and \qt{ML auditing tools [S10]}. Three students discussed learning about how AI absorbs \qt{bias} and \qt{unwanted content} as well as how to remove them. Remaining opinions varied, but some of these included navigating tradeoffs such as \qt{How the government / law enforcement leverages ML/AI in criminal prosecutions and how that meshes with personal privacies [S19]} and \qt{How to balance interaction between human and machine [S30]} as well as studying multi-step processes: \qt{I would like to learn about the development process for AI and how they determine the use/purpose of AI. [S35]}

\xhdr{Changes in perceptions of AI use cases} In response to prompting for additional use cases that they are excited about, several students mentioned communication on social media, and autonomous driving to control traffic. No further detail was provided. Regarding new concerning usages, at least six students expressed new concern about \texttt{generative models}: \qt{Just having heard about Vall-E today in the group activity it makes me a little uneasy knowing that there is a tool that easily [accessible] for someone to mimic my voice. [S29]} At least four participants described increased awareness of concerns about facial recognition software: \qt{I have always been worried about facial recognition but I had not realized that it was being used by places like entertainment venues until seeing two separate incidents on AIID today about lawyers being banned from Radio City Music Hall and MSG. [S39]} Three students raised concerns about autonomous driving. One reasoned \qt{I'm worried that the issues with self driving cars may persist until every car is self driving and can communicate with each other. [S31]} Medicine also came up three times, but only one student provided a reason: \qt{[I am still concerned about] ML in healthcare [...] especially because they have a tendency to have the largest impact in a person's life! [S12]} Three students attributed their newfound concerns to the widely accessible nature of AI technologies: \qt{ML models can be used by anyone, and as a result, someone with limited consideration of societal impacts can generate outputs harmful for others in society. [S30]}

\xhdr{Changes in beliefs around the role of ML experts} The majority of students continued to believe in the crucial role of ML experts and practitioners in addressing potential harms, but the AIID did not help them get a better sense of how this goal could be operationalized: \qt{I believe that MLEs can try to promote all of the above values by keeping the socioeconomic and sociopolitical impacts of ML models in mind when designing them, but from a technological standpoint I'm not sure how developers may be able to do so. [S19]}
One student doubted the will to do better: \qt{Honestly a lot of the stories I read about for the individual part of class dealt with stories of people being potentially at risk at being physically harmed and I think a lot of time many of the engineers knew the risk these programs had, but did not do anything to fix the issues until the media started raising issues. [S29]}
Several students expressed a newfound appreciation for accountability mechanisms:
\qt{In the incident I read, accountability or responsibility of the people who created the model was not clear. I felt making it clear is necessary for solving the issues. [S30]} Another said \qt{It struck me how few incidents had been truly resolved, so machine learning experts and engineers need to be accountable to addressing issues and maintaining self-governance and integrity. [S14]} One student mentioned the need for the right incentives to encourage responsible conduct: \qt{Building these protections for ML systems might seem like additional work, there should be regulatory incentives and periodic audits for companies/government entities using these systems. [S10]}

\vspace{-2mm}
\subsection{Areas of Improvement for AIID}\label{sec:AIID-feedback}

While many students found AIID useful, they overall had different experiences concerning the usability of the interface as well as recommendations for quality improvements. In our post-activity survey, most students noted that to them, the database was ``Easy'' or ``Very Easy'' to use, but approximately one-third of the respondents reported ``Neutral'' experiences or stated that the database was ``Hard'' or ``Very Hard'' to use.
Figure \ref{fig:usability-ratings} in Appendix \ref{app:usability} shows how all students rated the site's usability.
In open-ended responses, students stated that the tool needs to be promoted for \texttt{wider contribution \& use}. 
They also believed that \texttt{accelerating the review process} through (partial) automation integration with social media could additionally help AIID users.

\begin{figure}
    \centering
    \hspace{0.08\textwidth}
     \begin{subfigure}{0.45\textwidth}
         \centering
         \includegraphics[width=\textwidth]{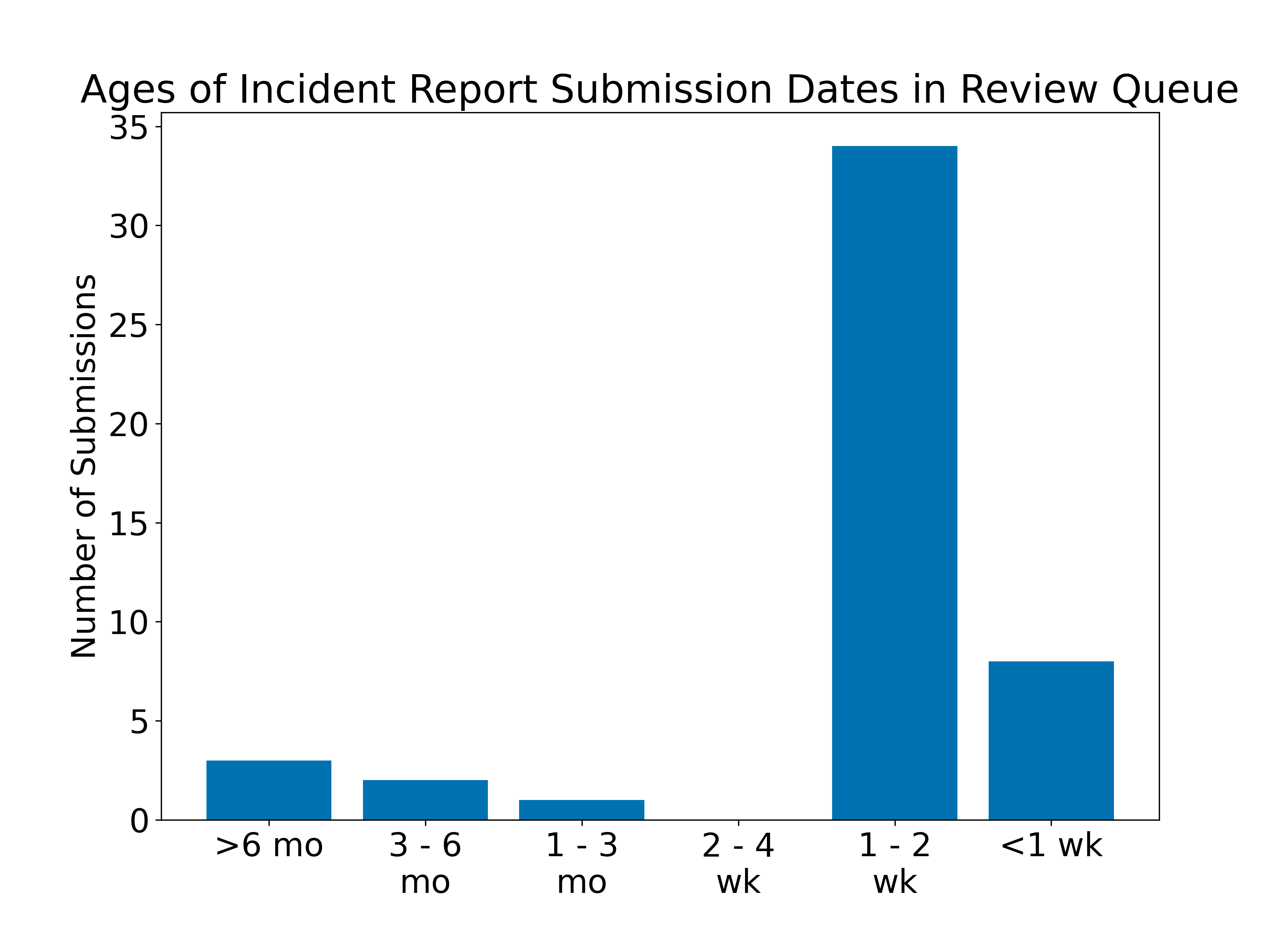}
     \end{subfigure}
     \hfill
     \begin{subfigure}{0.45\textwidth}
         \centering
         \includegraphics[width=\textwidth]{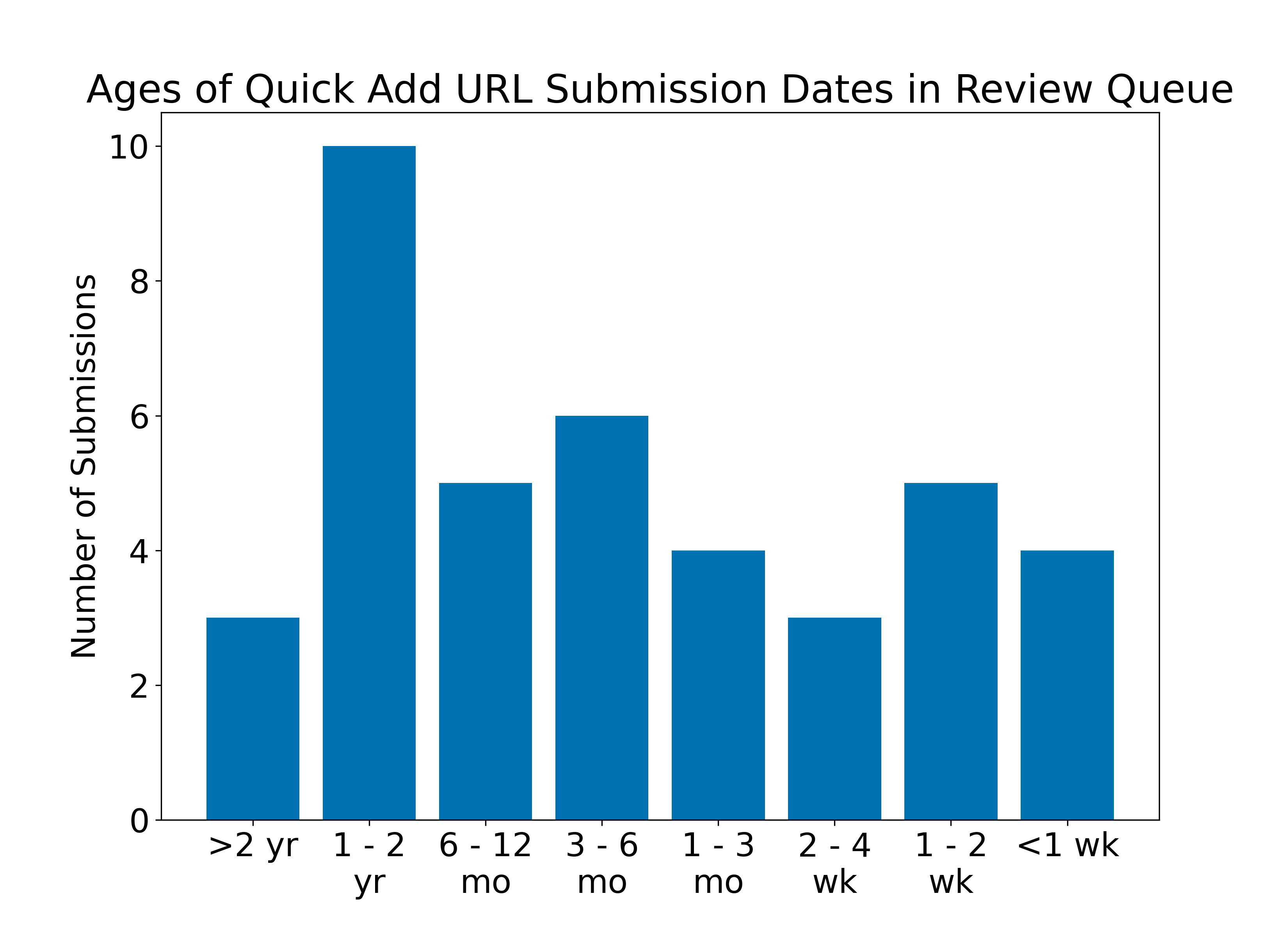}
     \end{subfigure}
     \hspace{0.08\textwidth}
     \vspace{-4mm}
     \caption{Age breakdowns of full incident report submissions and quick add URL submissions in the AIID review queue at the time of writing.}
     \label{fig:submission-ages}
\end{figure}

\xhdr{Follow-up AIID Analytics}
Nearly two weeks after the class activity, we inspected the database ourselves 
to study the report submission review queue 
and qualitatively analyze sources used in approved reports. To achieve the first, we downloaded the 
lists of incident reports and \emph{quick add URLs} (links to news articles about AI-related incidents submitted without a full report) in the review queue and 
bucketed submissions based on how long they have been in the queue.
The results are in Figure \ref{fig:submission-ages}. Through this exercise, we found that most full incident reports in the review queue are 1 - 2 weeks old. These reports include the ones submitted by our students, which had not been addressed by the time of this follow-up analysis (see Appendix \ref{app:queue}).
In comparison, \textit{quick add URL} submissions were more evenly distributed in terms of age in addition to being older in general. For instance, many of these latter submissions were 1 - 2 years old, but almost an equal number of submissions were under 1 month old. While this analysis only examines the queue at a particular point in time, it still supports the notion that speeding up the review process may help improve the recency and usefulness of the database, an idea echoed by students in their feedback.

\begin{SCfigure}
    \centering
    \includegraphics[width=0.4\textwidth]{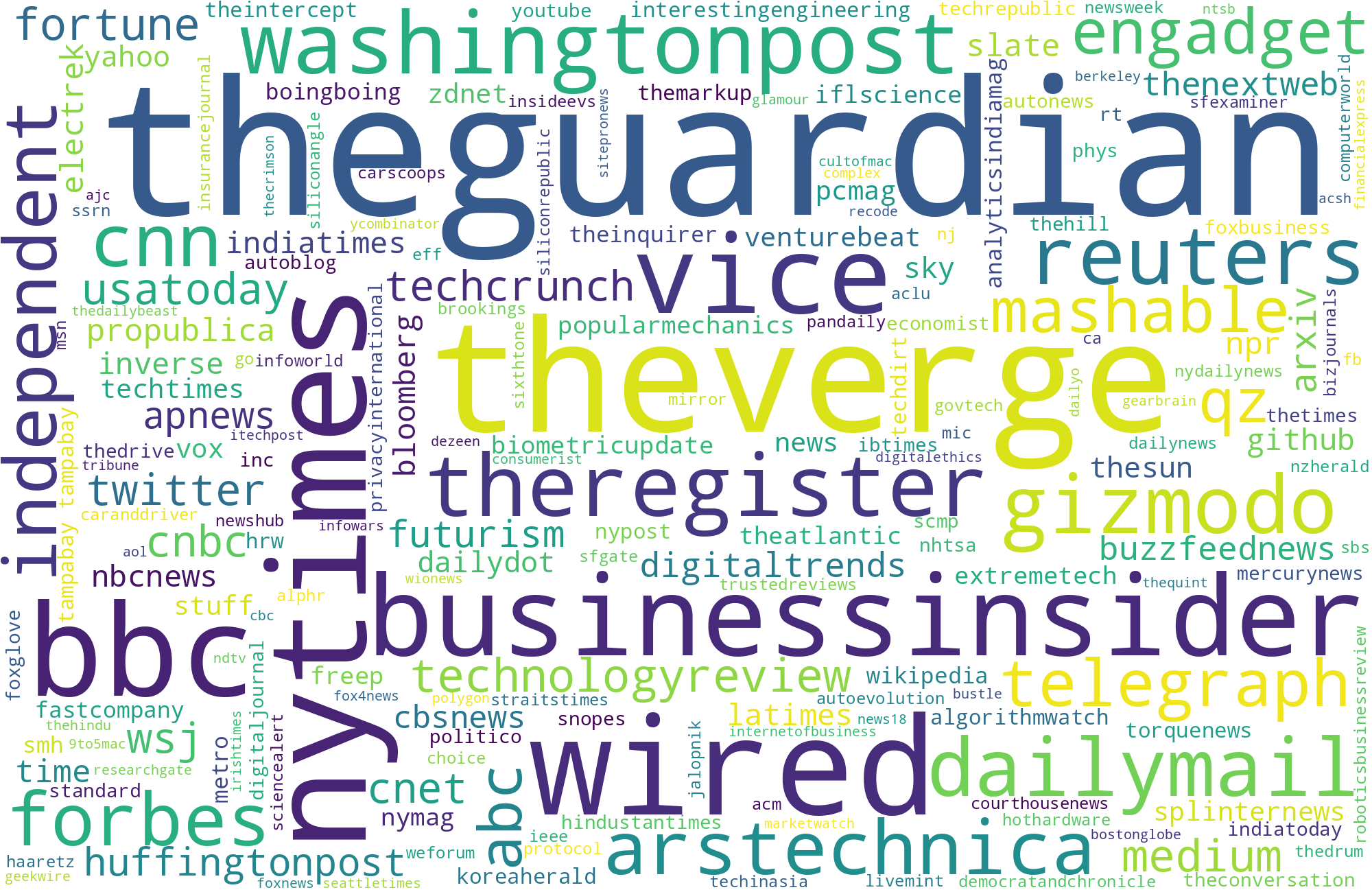}
    \caption{Word cloud of sources cited in incident reports listed on the AI Incident Database. Larger text corresponds to higher usage frequency in filing reports. A mixture of conventional journalism and tech-specific journalism are typically used as sources in incident reports. At the same time, independent blogging and reporting are also referenced as sources.}
    \label{fig:source-cloud}
    \vspace{-5mm}
\end{SCfigure}

To analyze the sources linked in approved incident reports, we used web scraping to obtain reports referenced in approximately 450 incidents in the database (almost all of the incidents contained in the database at the time of writing). We then processed the links to articles listed in these reports such that only the domain of each link was kept (e.g., ``blogs.wsj.com'' became ``wsj'', ``wired.co.uk'' became ``wired'', ``nytimes.com'' became ``nytimes'', etc.). Lastly, we generated a word cloud with the domains such that more frequently represented domains were rendered in a larger font than less frequently represented ones. The resulting image is in Figure \ref{fig:source-cloud}. A combination of typical news sources (such as the New York Times and BBC) and tech-specific reporting (such as Wired and Ars Technica) are predominant sources in these incident reports. Several independent journalism sources are also present. The presence of Medium and Twitter in the cloud suggests that posts beyond canonical reporting spheres are also referenced as sources in the database. This breadth of reporting may be why students found the database's information useful and at times surprising.

\vspace{-2mm}
\section{Discussion}\label{sec:discussion}
In this section, we discuss several limitations of our work, as well as its implications for the future development of AI incidents databases and AI ethics curricula. 

\xhdr{Limitations}
The primary limitation of our study is the small and selective sample of students from whom we gathered our data. First, our sample was composed of just under 40 students in total. Therefore, while we perform statistical tests in Appendix~\ref{app:tests} to measure the significance of opinion changes due to the introduction of AIID, we caution against generalizing these results due to our small sample size. Additionally, our participant pool was highly homogeneous (e.g., most had STEM backgrounds and were of Asian descent; all participants were students at a single R1 institution). With that in mind, our findings may not generalize to other student populations. Enrolling in the course and consenting to participate in this study may have introduced additional biases, such as self-selection, so our results do not directly apply to a random sample of CS students, even at our institution. Though such limitations are consistent with existing literature, we invite other educators and researchers to repeat this study in other environments, perhaps with modifications to the sampling procedure, to assess the extent to which the themes presented here generalize. 
In addition, while team and social interactions during the classroom activity allowed students to exchange ideas and findings regarding the AIID and ethics concepts, we note that they may have influenced students' post-activity responses alongside interaction with the database itself.  
Future work could examine the degree to which individual AIID exploration affects students' prioritization of RAI values.
Lastly, our method of analyzing the submission review queue may overestimate the review duration on AIID. Specifically, we studied the queue based on \emph{issues currently present in the queue} as opposed to \emph{all issues that have passed through the queue}. Note that the issues currently in the queue may be more difficult to review than the average issue, which could lead to biased estimates of turnaround time. To our knowledge, the database does not allow for comparing issue submission times with the corresponding issue approval times, so we could not come up with a more accurate estimate. 

\xhdr{Takeaways for AIID Development}
Our study suggests that AIID is a useful resource for raising awareness of AI harms to students, but there is room for improvement. First, considering the fact that our students were able to submit new reports found using their social media, one avenue of improvement consists of creating tools to integrate reporting with social media feeds, perhaps by making it possible to report incidents via a browser plugin and at the click of a button.
Another avenue of improvement concerns the speed and efficiency of the database. Students described how speeding up the review process might help AIID's credibility and usability. Our analysis of AIID's review queue supports this idea. Partial automation of certain review-related tasks (e.g., prioritizing reports for review or flagging duplicate reports) may support and speed up human oversight. (We recognize that human oversight and metadata curation adds significant value to the database. Thus, automation should only provide a supporting role.) Lastly,  
beyond leaderboards, alternative schemes to incentivize broader reporting (e.g., \emph{bias bounties} \cite{kenway2022bug,globus2022algorithmic}) could increase the usage and usefulness of the database over time.

\xhdr{Takeaways for AI Ethics Education}
Our qualitative findings around student educational motivations and goals corroborate recent reports of the dire need for AI ethics to become a core and standard part of AI curricula~\citep{stavrakakis2021teaching,fiesler2020we}: taking our institution as an example, while there are a wide range of technology ethics and societal impacts courses offered to students, the majority of these courses are inconsistently offered electives that focus on special (mostly computational) topics.
Students' desire for building awareness and developing their mental ``radar'' for ethical considerations, especially those that they may currently be unaware of, suggests AI education should help raise students' awareness so that they can avoid blind spots in the future \citep{madaio2020co}.
Additional learning objectives that our work recommends for inclusion in AI ethics curricula are (a) offering a systematic conceptual framework to think about common AI ethics issues; (b) supporting students' efforts to improve their critical thinking skills; and (c) building capacity for constructive dialog with wide ranges of AI stakeholders.

Our analysis suggests that the AIID can serve as a way to notify students of concepts and harms that they may not have originally considered, despite their backgrounds in ML/AI.
Additionally, the shift in what students considered more important after interacting with the AIID suggests that AI ethics education should emphasize safety and accountability relative to how they are currently covered alongside topics such as fairness, transparency, and privacy. This is in line with findings from \citet{raji2022fallacy} who note that even those who wish to regulate AI, such as by reducing bias or minimizing harm, assume that AI is functional when this may not necessarily be the case. It is important to bring this awareness of safety and accountability issues into educational settings so that future developers and practitioners have the requisite tools and knowledge to properly handle problems in this space.
Performing failure case analyses is consistent with other systems engineering ethics education \cite{colby2008ethics} including in software engineering \cite{leveson1993investigation,huff2004integrating}. 
We believe the AIID can be a rich source for finding AI incident case studies for use in the classroom.
\vspace{-2mm}
\section{Conclusion}
In this study, we explored the use of the AI Incident Database as an educational tool to help raise students' awareness of AI harms.
Through a classroom activity exposing students to the AIID and a set of pre- and post-activity surveys, we discovered that students found the database useful in exposing them to never-before-seen AI incidents and this exposure may have shifted their thinking toward AI \textit{safety} and \textit{accountability}.
Although limited to one class at a single institution, we believe that our results suggest the value of AIID within AI Ethics education and can serve to help improve students' awareness of potential AI harms.
We hope to see expanded use of the database and further improvements to its user base and usability.
We aspire to move the needle in helping future AI practitioners consider the risks of AI implementations and overcome failures of imagination by encouraging exposure to incident reports early in their careers. Our aim is that this contribution, when combined with other carefully designed educational strategies, will result in safer AI systems and more accountable AI-building organizations.

\section*{Acknowledgements}
\mfedit{
H. Heidari acknowledges support from NSF (IIS2040929 and IIS2229881) and PwC (through the Digital Transformation
and Innovation Center at CMU). M. Feffer acknowledges support from the National GEM Consortium and the ARCS Foundation. Any opinions, findings, conclusions, or recommendations expressed in this material are those of the authors and do not reflect the views of the National Science Foundation and other funding agencies.}

\pagebreak
\bibliographystyle{apa-good}
\bibliography{main_FAccT.bib}

\pagebreak
\appendix
\section{Consent and Data Collection Processes}\label{app:consent}

To obtain informed consent and anonymize students' data, the requisite processes started by the course instructor emailing the following:

\begin{quote}
    Dear all,

    As I have mentioned in the course syllabus and in prior lectures, my team and I plan to conduct a research study assessing the effectiveness of several educational tools and resources I will utilize in this course during the semester. We are writing to invite you to participate in this study.

    You will not be asked to do anything beyond the normal activities and assignments that are part of the course curriculum. You have the choice of opting out of data collection for research purposes. Participants will not receive any compensation, and those opting out will not not have their course outcome adversely impacted in any way.

    Please see the opt-out form attached. If you wish to opt out of the study, please sign the form indicating that and send it to [student advised by instructor but not part of teaching staff] preferably before the class on [date of first research activity]. \textbf{Please do NOT cc the course staff. This way the teaching staff (including me) won’t know whether you opted out until after all grades have been posted and submitted. This is to ensure that you won’t feel coerced to participate in our study.}

    If you have any questions or concerns at any point, please don’t hesitate to get in touch with me. 

    Best regards,
    
    [Instructor] 
\end{quote}

The opt-out consent form contained the following information regarding data collection and usage: 

\begin{quote}
    Data collection: (1) In the classroom, you (the study participant)  will complete a brief pre-study questionnaire aiming to assess your initial understanding of the concepts. (2) You will then learn about the concepts and the corresponding tool through the instructor. (3) You will be asked to work in teams with the tool and submit a short report in collaboration with your teammates. If all students in your team consent, we will ask you to record the audio of your deliberation via [university's] Zoom rooms and share the recording or its transcript with us. (4) You will complete a survey and participate in a class-wide evaluation of the activity. (5) Finally, one week after the class activity, you will complete a post-activity questionnaire/quiz designed to assess your final understanding of the concepts.

    Most questions in our questionnaires will require open-ended responses. Please do not reveal any private or personally-identifiable information about yourselves or others in your answers to the open-ended questions. We will also ask you to provide us with basic information about your educational background and demographics. The purpose of these questions is for us to detect any significant variations in responses across the corresponding dimensions. Answering demographic questions is entirely optional. 
\end{quote}

In accordance with this outlined protocol, a Python script was run to anonymize all data by converting all email addresses and other identifiable information to random IDs in the form of integers starting from 1. No members of the research team that were also teaching staff observed identifiable information prior to running this script.

\section{Pre-class Questionnaire (Verbatim)}\label{app:pre-class-questionnaire}

This questionnaire aims to understand your background, knowledge, and goals as they relate to Machine Learning (ML), Artificial Intelligence (AI), and the topics covered in [...] (our class). (Please respond to the best of your knowledge and memory and don't consult any external resources. There is no right or wrong answer to any of these questions). It also asks an optional set of questions about your demographics. Please feel free to leave that last part un-answered. 

Your answers to this questionnaire will:
    \begin{enumerate}
        \item inform the themes, topics, and tools we will focus on during the semester.
        \item If you consent, a de-identified version of your collective responses will be utilized in a research study by [...] (the instructor), in which she aims to assess the effectiveness of the educational tools she utilizes in this course. The opt-out consent form for the study can be found here. Please follow the instructions there if you wish to opt out.
    \end{enumerate}

We expect the questionnaire will take 10-15 minutes to complete (on average).

\subsection{Educational Background}
\begin{enumerate}
    \item Which one best describes your educational background?
    \begin{enumerate}[label=\Alph*.]
        \item Science, Technology, Engineering, Mathematics (STEM)
        \item Humanities, Social Sciences and the Arts (HSA)
        \item Other\ldots (Short Answer)
    \end{enumerate}
    \item What is your current major (i.e., the name of the degree program you are enrolled in)? (Short Answer)
    \item What is the highest degree or level of school you have completed?
    \begin{enumerate}[label=\Alph*.]
        \item High school or equivalent
        \item Bachelor degree or equivalent
        \item Master degree or equivalent
        \item Doctoral degree or equivalent
        \item Other\ldots (Short Answer)
    \end{enumerate}
    \item Have you completed any courses with a \underline{significant ML component} in the past? If yes, please list those courses. (Short Answer)
    \item Which one best describes your level of familiarity with AI/ML?
    \begin{enumerate}[label=\Alph*.]
        \item None
        \item Elementary
        \item Intermediate
        \item Advanced
        \item Other\ldots (Short Answer)
    \end{enumerate}
    \item Have you received any form of training on \underline{ethical and societal considerations} around the use of AI or ML?
    \begin{enumerate}[label=\Alph*.]
        \item Yes
        \item No
    \end{enumerate}
    \item If your response to the previous question was "yes", please briefly describe the nature of the training. (Short Answer)
    \item Please briefly describe what motivated you to take this course. (Long Answer)
    \item What do you hope to learn from/gain out of this course? (Long Answer)
    \item Are there any specific \underline{\textit{topics, tools, or applications}} you would like to learn about as a part of this course?
\end{enumerate}

\subsection{ML in Society}
\begin{enumerate}
    \item Please name a use case/application of ML in society that you are \underline{excited} about. (Short Answer)
    \item How did you learn about this use case? (Short Answer)
    \item Please name a use case/application of ML in society that you are \underline{concerned} about. (Short Answer)
    \item How did you learn about this use case? (Short Answer)
    \item Which one of the following broad categories of \underline{values} do you believe is \underline{most urgent to address} to promote the responsible use of ML in socially high-stakes domains? (A high-level description of each value--taken from prior work--is provided below. Options are ordered randomly).
    
\begin{table}
    \caption{Students were offered a brief description of values taken verbatim from prior work of ~\citet{jakesch2022different}}
    \label{tab:value-descriptions}
{\footnotesize
\begin{tabular}{p{0.15\linewidth}|p{0.8\linewidth}}
    \textbf{RAI value} & \textbf{Description} \\
    \hline
    Transparency & A transparent AI system produces decisions that people can understand. Developers of transparent AI systems ensure, as far as possible, that users can get insight into why and how a system made a decision or inference. \\
    \hline
    Fairness & A fair AI system treats all people equally. Developers of fair AI systems ensure, as far as possible, that the system does not reinforce biases or stereotypes. A fair system works equally well for everyone independent of their race, gender, sexual orientation, and ability \\
    \hline
    Safety & A safe AI system performs reliably and safely. Developers of safe AI systems implement strong safety measures. They anticipate and mitigate, as far as possible, physical, emotional, and psychological harms that the system might cause. \\
    \hline
    Accountability & An accountable AI system has clear attributions of responsibilities and liability. Developers and operators of accountable AI systems are, as far as possible, held responsible for their impacts. An accountable system also implements mechanisms for appeal and recourse. \\
    \hline
    Privacy & An AI system that respects people’s privacy implements strong privacy safeguards. Developers of privacy-preserving AI systems minimize, as far as possible, the collection of sensitive data and ensure that the AI system provides notice and asks for consent. \\
    \hline
    Autonomy & An AI system that respects people’s autonomy avoids reducing their agency. Developers of autonomy-preserving AI systems ensure, as far as possible, that the system provides choices to people and preserves or increases their control over their lives. \\
    \hline
    Performance & A high-performing AI system consistently produces good predictions, inferences or answers. Developers of high-performing AI systems ensure, as far as possible, that the system’s results are useful, accurate and produced with minimal delay. \\
    \hline 
    \end{tabular}
}
\end{table}

    \begin{enumerate}[label=\Alph*.]
        \item Fairness
        \item Safety
        \item Transparency
        \item Privacy
        \item Accountability \& governance
        \item Human autonomy \& agency
        \item Performance \& efficiency
        \item Other\ldots (Short Answer)
    \end{enumerate}
    \item How do you believe \underline{\textit{Machine Learning experts and engineers}} can contribute to promoting the above value? Please elaborate. (Long Answer)
\end{enumerate}

\subsection{AI News}
\begin{enumerate}
    \item How often do you normally hear about AI-related news (e.g., new use cases, significant advances, etc.)?
    \begin{enumerate}[label=\Alph*.]
        \item Every day
        \item Every week
        \item Every month
        \item Every year
        \item Rarely
    \end{enumerate}
    \item Where (or from what source) do you usually hear about AI-related news? (Options below are ordered randomly.)
    \begin{enumerate}[label=\Alph*.]
        \item Educational resources (e.g., lectures, course material, university mailing lists, \ldots)
        \item Social media (e.g., Twitter, Facebook, LinkedIn \ldots)
        \item Traditional media (e.g., TV, Radio, printed or online newspapers, \ldots)
        \item Word of mouth (e.g., through classmates, friends, and acquaintances)
        \item Other\ldots (Short Answer)
    \end{enumerate}
    \item Please briefly describe an example of recent AI news you have heard about. (Long Answer)
\end{enumerate}

\subsection{Demographic Information (Optional)}
The purpose of this section is to understand whether there are significant variations in your responses to the previous questions along demographic dimensions. Answering this part is entirely optional, so please feel free to leave a question unanswered if you prefer not to disclose the corresponding information about yourself.  All categorical alternatives in this section have been ordered alphabetically.

\begin{enumerate}
    \item Which one better describes your political views – please pick the closest one if neither is an exact fit?
    \begin{enumerate}[label=\Alph*.]
        \item Conservative
        \item Liberal
        \item Libertarian
        \item Other\ldots (Short Answer)
    \end{enumerate}
    \item Which one best describes your gender?
    \begin{enumerate}[label=\Alph*.]
        \item Female
        \item Male
        \item Non-binary
        \item Other\ldots (Short Answer)
    \end{enumerate}
    \item What is your age group?
    \begin{enumerate}[label=\Alph*.]
        \item 18-26
        \item 27-40
        \item 41 or older
        \item Other\ldots (Short Answer)
    \end{enumerate}
    \item Which one best describes your race?
    \begin{enumerate}[label=\Alph*.]
        \item Asian
        \item Black or African American
        \item Native Hawaiian or Other Pacific Islander
        \item White
        \item American Indian or Alaskan Native
        \item Other\ldots (Short Answer)
    \end{enumerate}
    \item Do you believe you belong to a marginalized/disadvantaged group or community?
    \begin{enumerate}[label=\Alph*.]
        \item Yes
        \item No
        \item Other\ldots (Short Answer)
    \end{enumerate}
\end{enumerate}

\subsection{Conclusion}
Thank you so much for your participation!

\begin{enumerate}
    \item If you have any feedback for the teaching instructor or the research team about the questionnaire, please leave your comments here. (Long Answer)
\end{enumerate}

\section{In-class Activity}\label{app:in-class-activity}

\subsection{Individual activity}
\begin{enumerate}
    \item Go to “Table View” tab, the search page corresponding to your number.
    \item Read the summary of the 10 incidents in your page.
    \item Among the 10, pick the incident that grabs your attention (e.g., because it’s news to you, it’s surprising, the magnitude of impact could be significant,  … )
    \item Read the incident report carefully.
    \item Then provide 1--2 word responses to the next questions.
    \begin{itemize}
        \item What was the \underline{source} of the story (e.g., which website/author published it)?
        \item In what \underline{application domain} did the incident occur?
        \item What was the \underline{nature of the incident} (i.e., the concern that was raised)?
        \item Who was (partially) \underline{responsible} for the incident (e.g., because they developed, deployed, or used the AI system)?
        \item Who was (potentially) \underline{harmed}?
        \item How was the incident ultimately \underline{addressed} (Was there a penalty? Was the tool discontinued?)
    \end{itemize}
\end{enumerate}

\subsection{Team activity}
\begin{enumerate}
    \item Form a team with the 4 classmates sitting closest to you.
    \item Search for a recent AI/ML incident that has not been submitted to the database.
    \begin{itemize}
        \item If you don’t know where to start, go back to the AIID and look at alternative views of the data (e.g., spatial, entities, …) or the review queue for inspiration. 
        \item It is okay if you can’t find such a story after looking for 15 minutes.
    \end{itemize}
\item Produce a team report as follows and upload it to the course website individually.
    \begin{itemize}
        \item Submit a report on your story to AIID. Take a snapshot of your report.
        \item If you didn’t find a story, briefly describe your search process (e.g., queries, websites).
    \end{itemize}
\end{enumerate}

\section{Post-Activity Questionnaire (Verbatim)}\label{app:post-activity-questionnaire}

This questionnaire aims to assess the efficacy of the first class activity in which we explored the AI Incident Database (AIID) by eliciting your feedback about the tool. As before, there is no right or wrong answer to any of these questions.

If you consent, a de-identified version of your collective responses will be utilized in a research study by [...] (the instructor), in which she aims to assess the effectiveness of the educational tools she utilizes in this course. The opt-out consent form for the study can be found here. Please follow the instructions there if you wish to opt out.

We expect the questionnaire will take around 7 minutes to complete (on average).

\subsection{Assessing the impact of AIID}
In the previous questionnaire, we asked you about 
    \begin{itemize}
        \item your motivation for taking this course and the topics you'd like to learn about,
        \item the applications of ML in society that you are excited/concerned about, 
        \item the values you believe should be prioritized, 
        \item and the role of ML experts in promoting those values. 
    \end{itemize}

The goal of the following questions is to understand whether working with AIID has impacted your response to any of the above questions.

\begin{enumerate}
    \item Having explored AIID, has your motivation for taking this course changed at all?
    \begin{enumerate}[label=\Alph*.]
        \item It hasn't changed.
        \item It has increased.
        \item It has decreased.
        \item Other\ldots (Short Answer)
    \end{enumerate}
    \item Please briefly describe why you selected the answer above. (Long Answer)
    \item Having explored the AIID, are there any \underline{new/additional} uses/applications of ML in society that you are now \underline{excited} about? (Long Answer)
    \item Having explored the AIID, are there any \underline{new/additional} uses/applications of ML in society that you are now \underline{concerned} about? (Long Answer)
    \item Having explored the AIID, which one of the following broad categories of values do you now believe is most urgent to address to promote the responsible use of ML in socially high-stakes domains? (A brief description of each value--taken from prior work--is provided below. Options are ordered randomly).

    \begin{enumerate}[label=\Alph*.]
        \item Fairness
        \item Safety
        \item Transparency
        \item Privacy
        \item Accountability \& governance
        \item Human autonomy \& agency
        \item Performance \& efficiency
        \item Other\ldots (Short Answer)
    \end{enumerate}
    \item Has exploring the AIID changed your belief about how \underline{\textit{Machine Learning experts and engineers}} can contribute to promoting the above value? Please elaborate. (Long Answer)
    \item Having explored the AIID, are there any additional tools or topics you would like to learn about in this course? (Long Answer)
\end{enumerate}

\subsection{Feedback on AIID}

\begin{enumerate}
    \item How likely are you to use AIID in the future?
    \begin{enumerate}[label=\Alph*.]
        \item I will refer to it frequently (i.e., at least every 1-2 weeks).
        \item I will refer to it occasionally (i.e., every few months).
        \item I will refer to it only if the need arises.
        \item I will likely never use it again.
        \item Other\ldots (Short Answer)
    \end{enumerate}
    \item Do you have any additional thoughts or comments on the \underline{limitations} of AIID? (Long Answer)
    \item How easy/difficult did you find it to work with the user interface of AIID?
    \begin{enumerate}
        \item Very Easy
        \item Easy
        \item Neutral
        \item Hard
        \item Very Hard
    \end{enumerate}
    \item Do you have any suggestions or ideas regarding how AIID can be \underline{improved}? (Long Answer)
\end{enumerate}

\subsection{Conclusion}
Thank you so much for your completing this assignment!

\begin{enumerate}
    \item If you have any feedback for the teaching instructor or the research team about this questionnaire in particular or the first session in general, please leave your comments here. (Long Answer)
\end{enumerate}

\section{In-class Findings}\label{app:in-class-findings}

\begin{figure}[h]
    \centering
    \begin{subfigure}{0.45\textwidth}
        \centering
        \includegraphics[width=\textwidth]{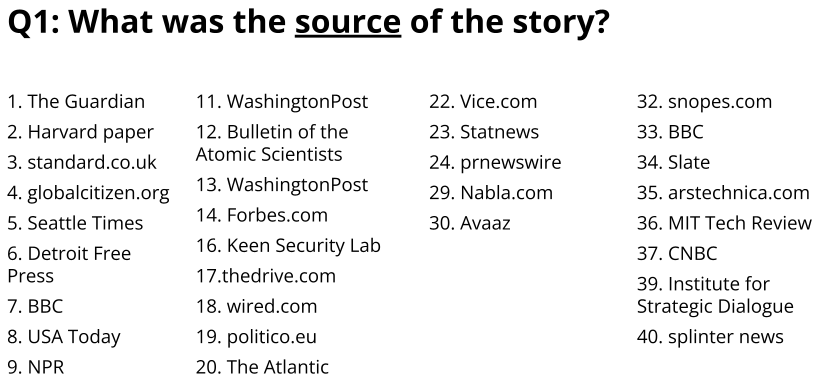}
    \end{subfigure}
    \hfill
    \begin{subfigure}{0.45\textwidth}
        \centering
        \includegraphics[width=\textwidth]{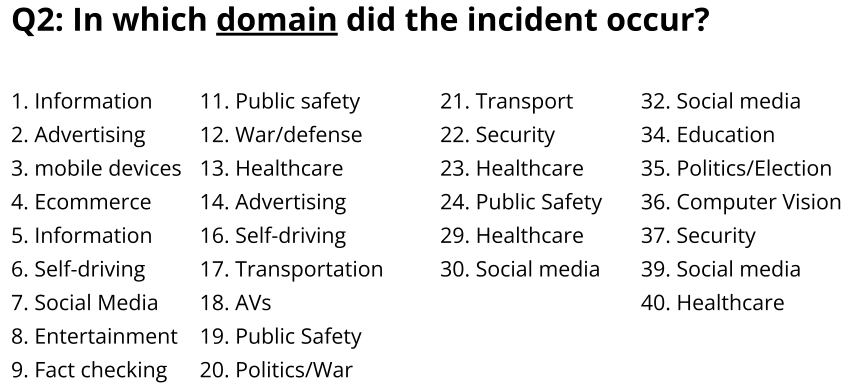}
    \end{subfigure}
    \\
    \begin{subfigure}{0.45\textwidth}
        \centering
        \includegraphics[width=\textwidth]{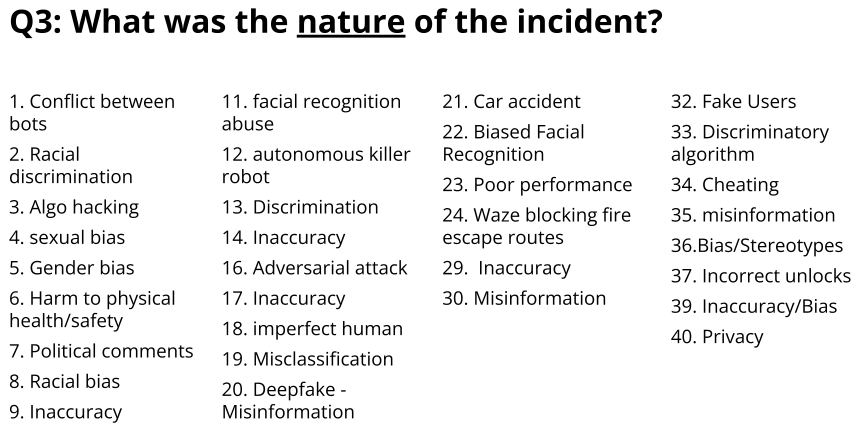}
    \end{subfigure}
    \hfill
    \begin{subfigure}{0.45\textwidth}
        \centering
        \includegraphics[width=\textwidth]{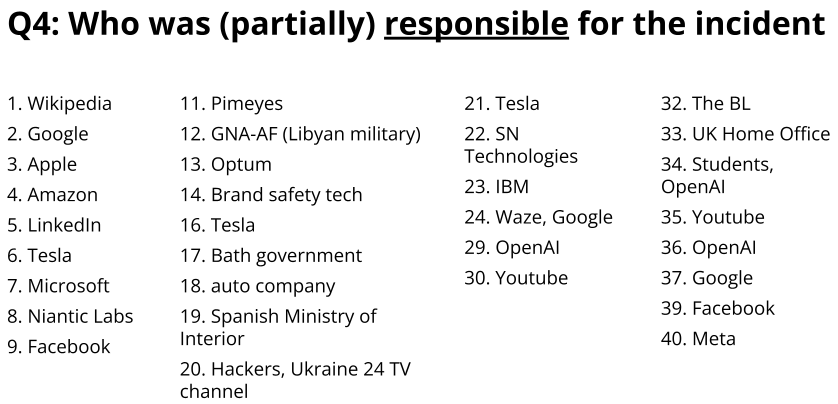}
    \end{subfigure}
    \\
    \begin{subfigure}{0.45\textwidth}
        \centering
        \includegraphics[width=\textwidth]{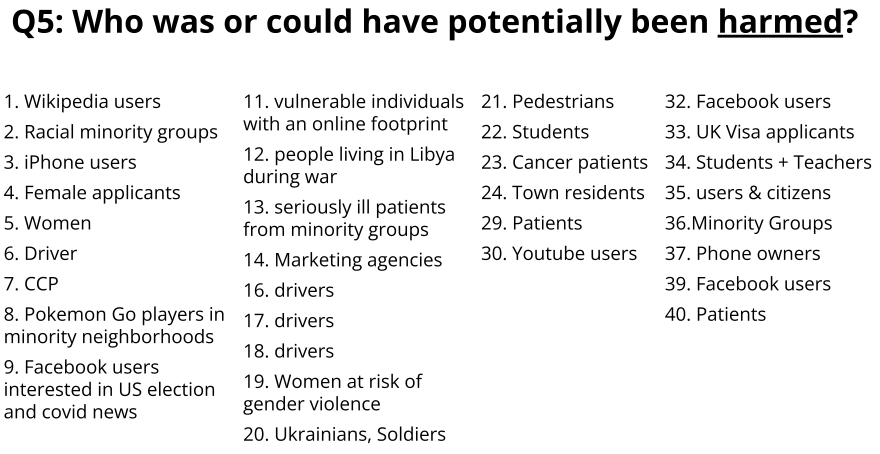}
    \end{subfigure}
    \hfill
    \begin{subfigure}{0.45\textwidth}
        \centering
        \includegraphics[width=\textwidth]{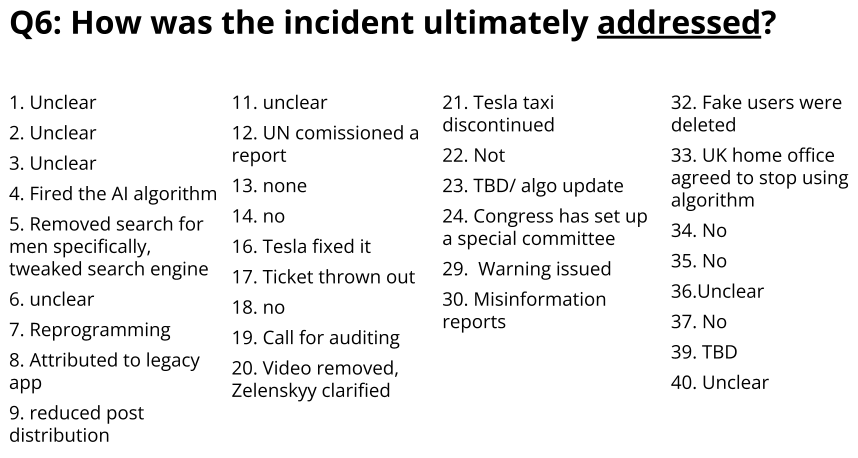}
    \end{subfigure}
    \caption{Results of students' explorations of AIID.}
    \label{fig:in-class-findings}
\end{figure}

\begin{figure}[h]
    \centering
    \includegraphics[width=0.5\textwidth]{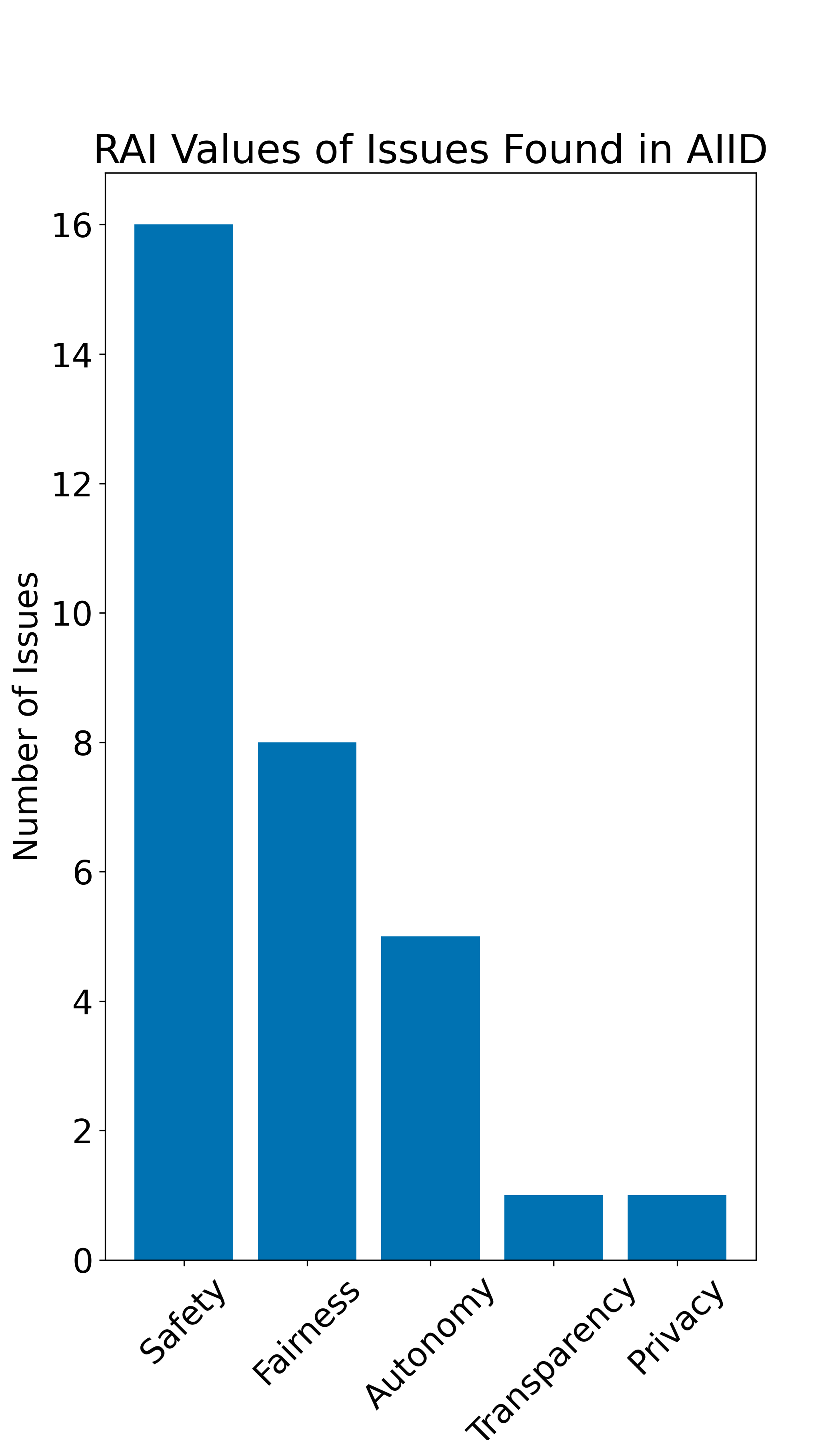}
    \caption{Value distribution of issues found interesting by students. Of the 31 issues explored, we identified
    16 as pertaining to the RAI value of Safety as per \citet{jakesch2022different}, 8 as pertaining to Fairness,
    5 as pertaining to Autonomy,
    1 as pertaining to Privacy, and 1 as pertaining to Transparency. We additionally note that Accountability \& Governance cannot be a reason for an incident of harm as calls for oversight stem from the results of harm.}
    \label{fig:issue-breakdown}
\end{figure}

\section{Participant Demographics}\label{app:demographics}
 
\begin{figure}[H]
    \centering
     \begin{subfigure}{0.2\textwidth}
         \centering
         \includegraphics[width=\textwidth,valign=t]{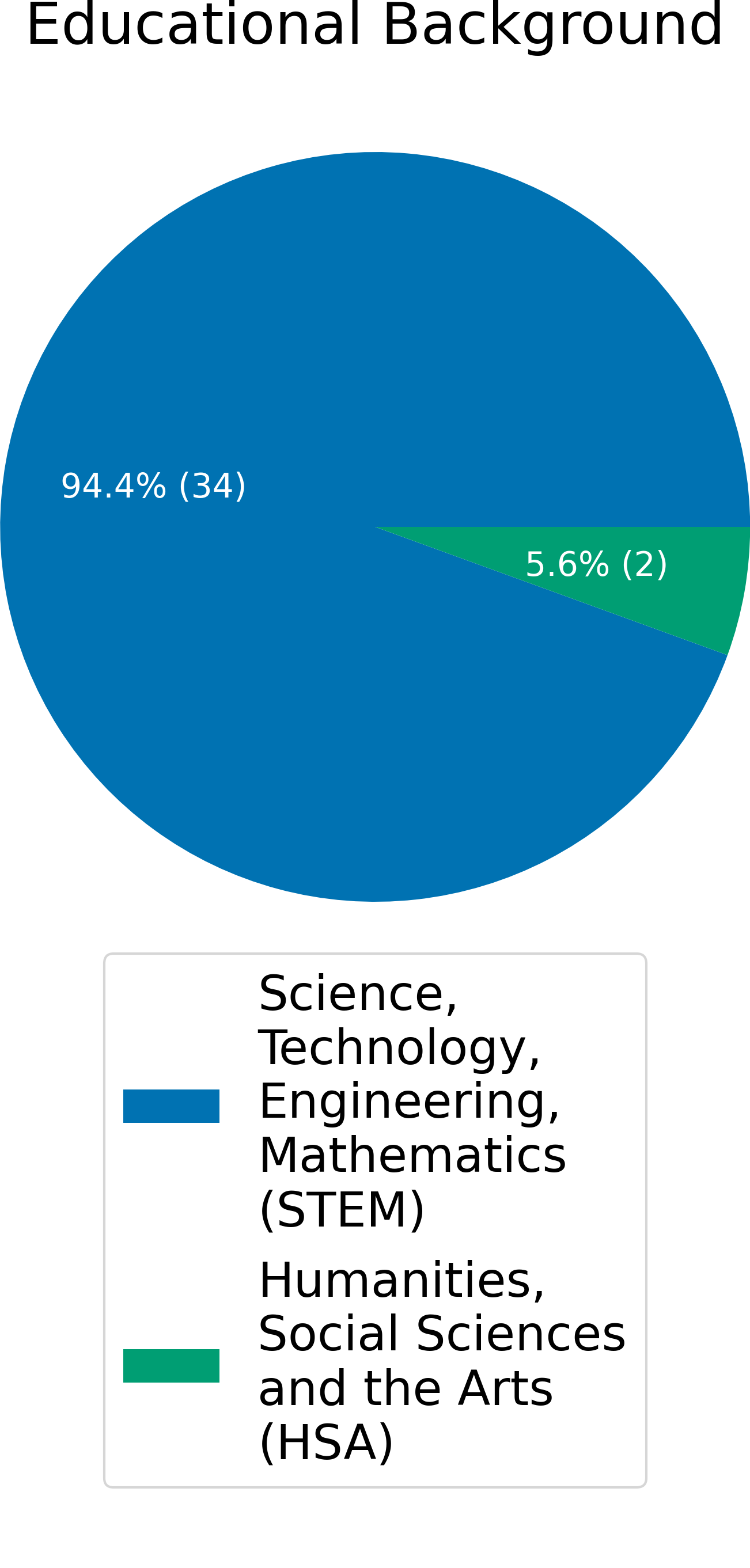}
     \end{subfigure}
     \hfill
     \begin{subfigure}{0.2\textwidth}
         \centering
         \includegraphics[width=\textwidth,valign=t]{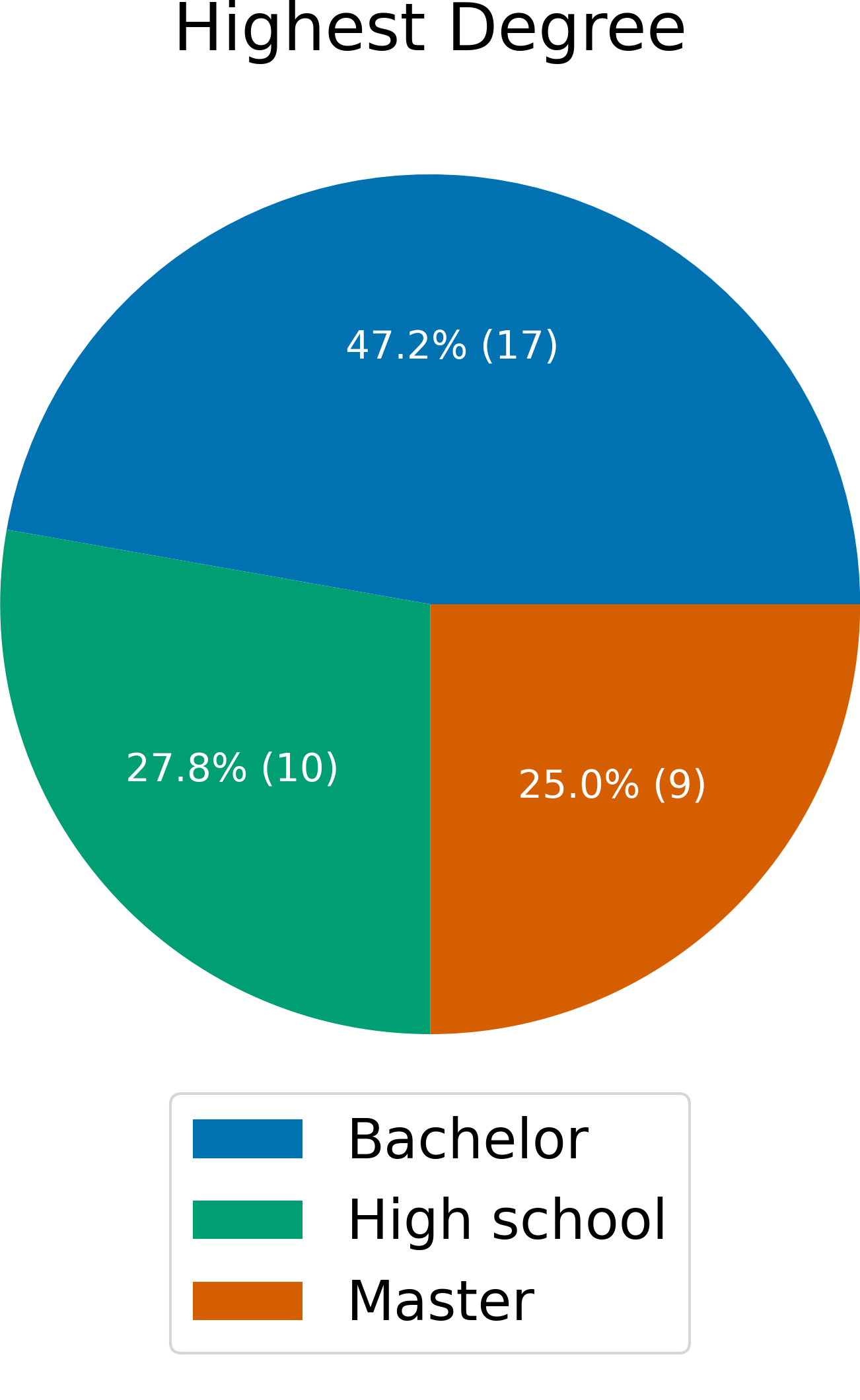}
     \end{subfigure}
     \hfill
     \begin{subfigure}{0.2\textwidth}
         \centering
         \includegraphics[width=\textwidth,valign=t]{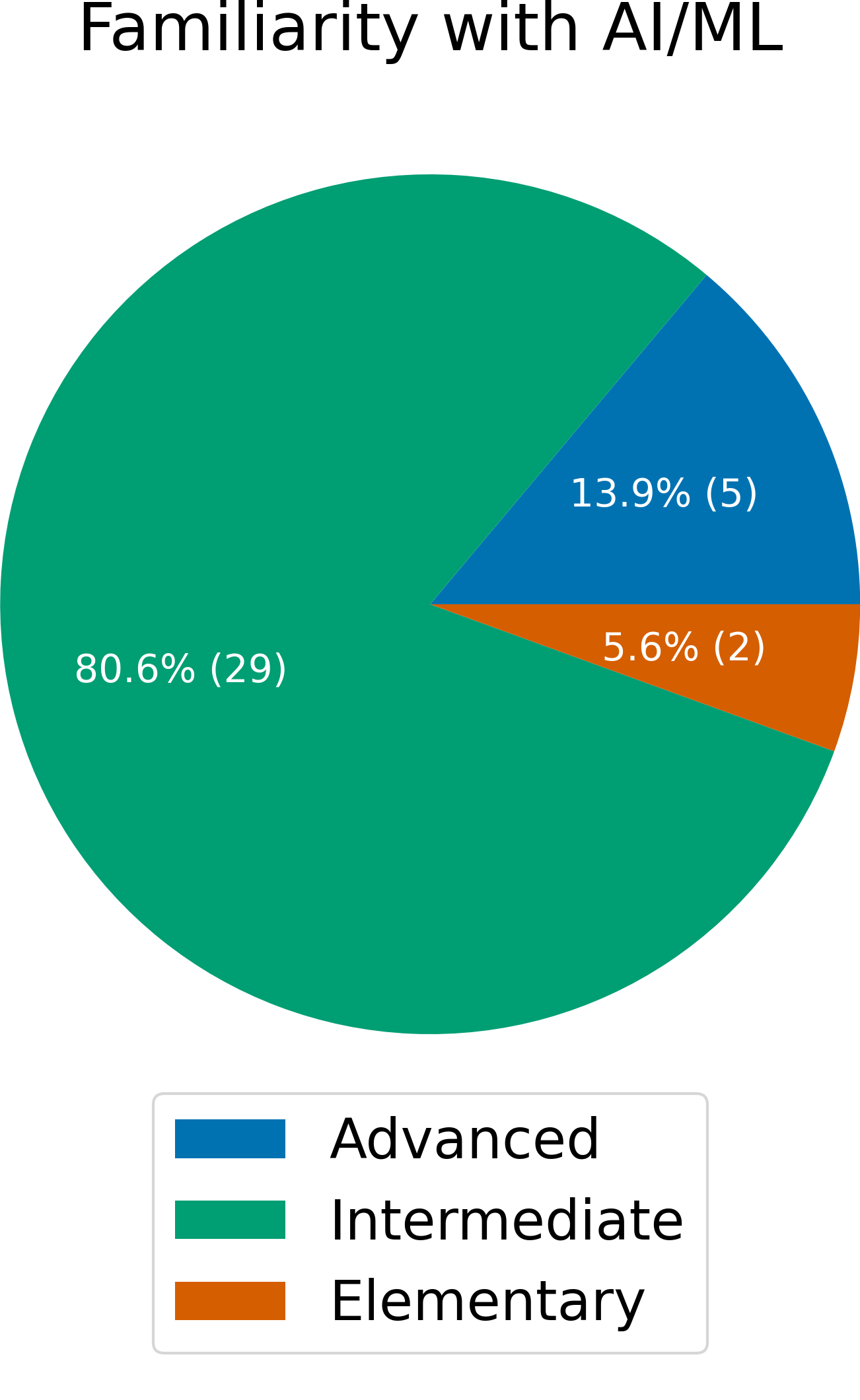}
     \end{subfigure}
     \hfill
     \begin{subfigure}{0.2\textwidth}
         \centering
         \includegraphics[width=\textwidth,valign=t]{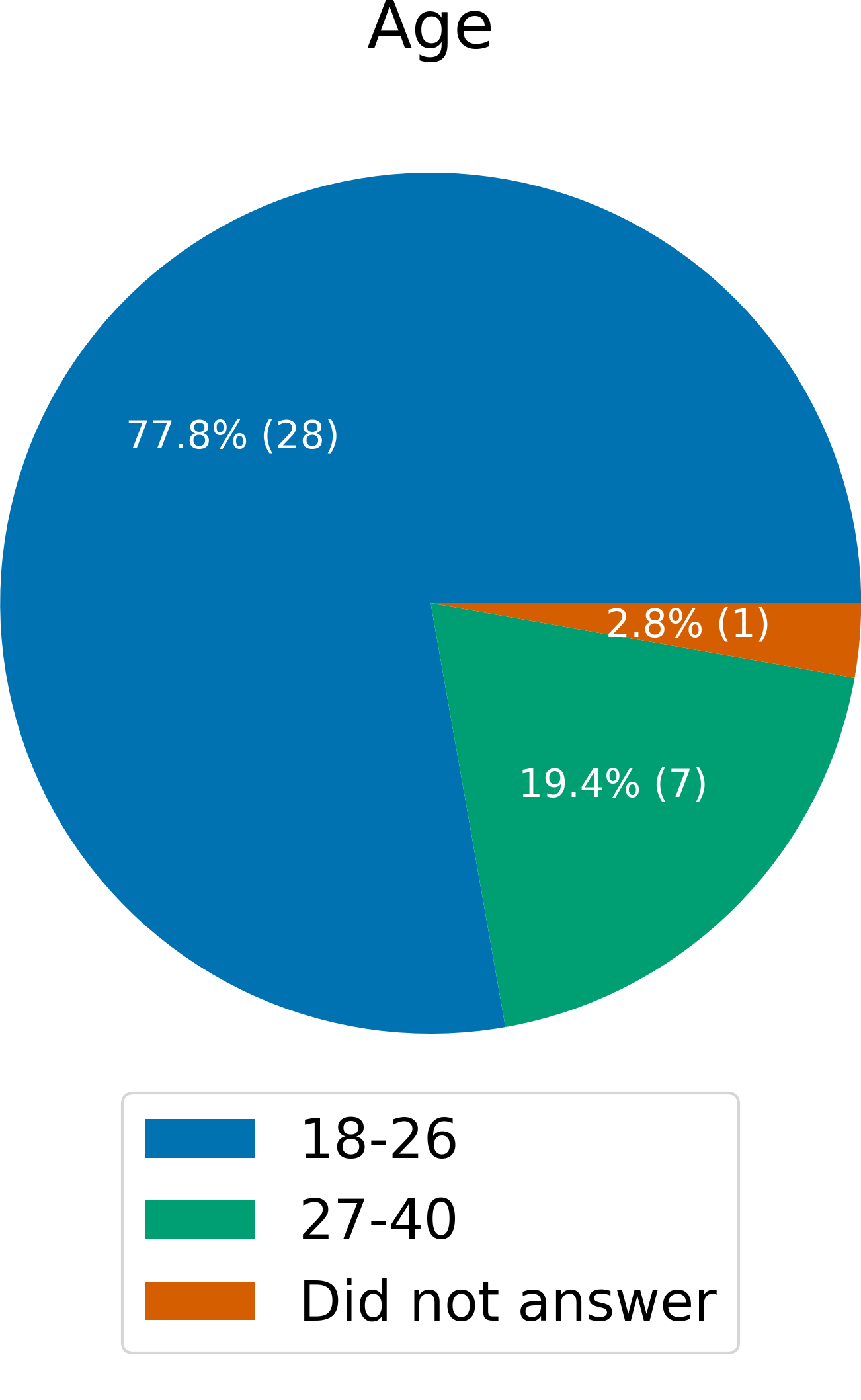}
     \end{subfigure}
     \\
     \begin{subfigure}{0.2\textwidth}
         \centering
         \includegraphics[width=\textwidth,valign=t]{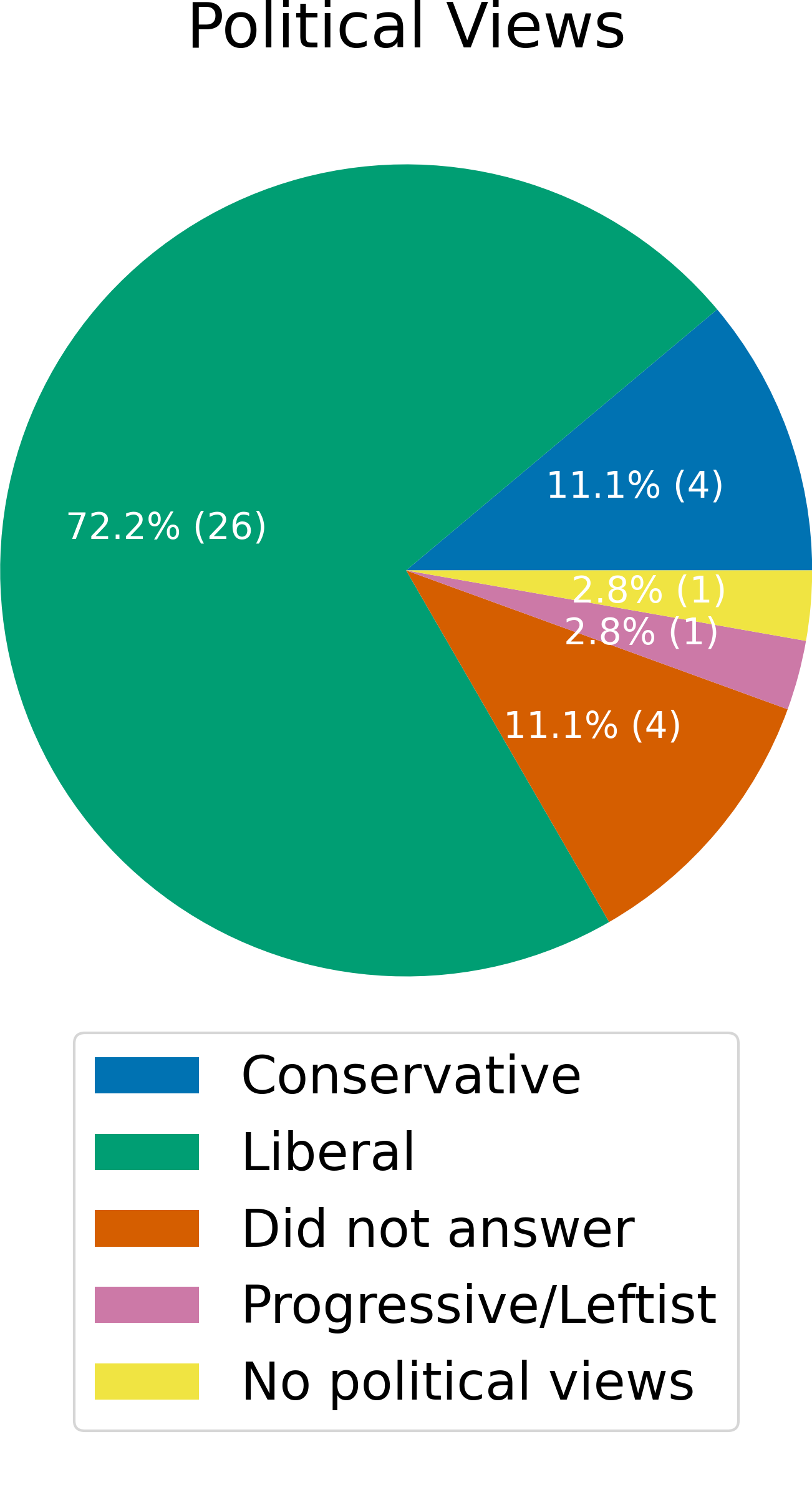}
     \end{subfigure}
     \hfill
     \begin{subfigure}{0.2\textwidth}
         \centering
         \includegraphics[width=\textwidth,valign=t]{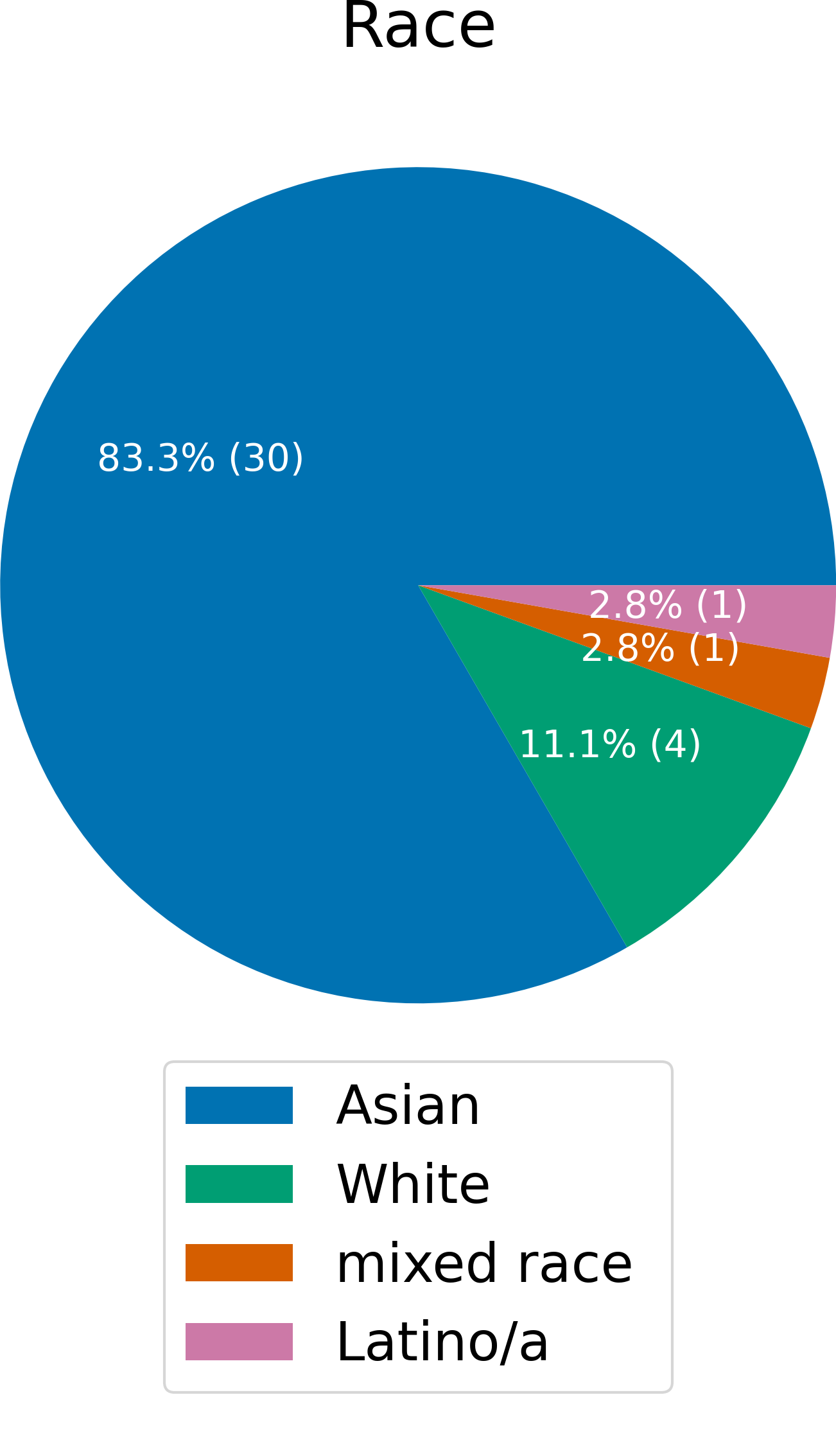}
     \end{subfigure}
     \hfill
     \begin{subfigure}{0.2\textwidth}
         \centering
         \includegraphics[width=\textwidth,valign=t]{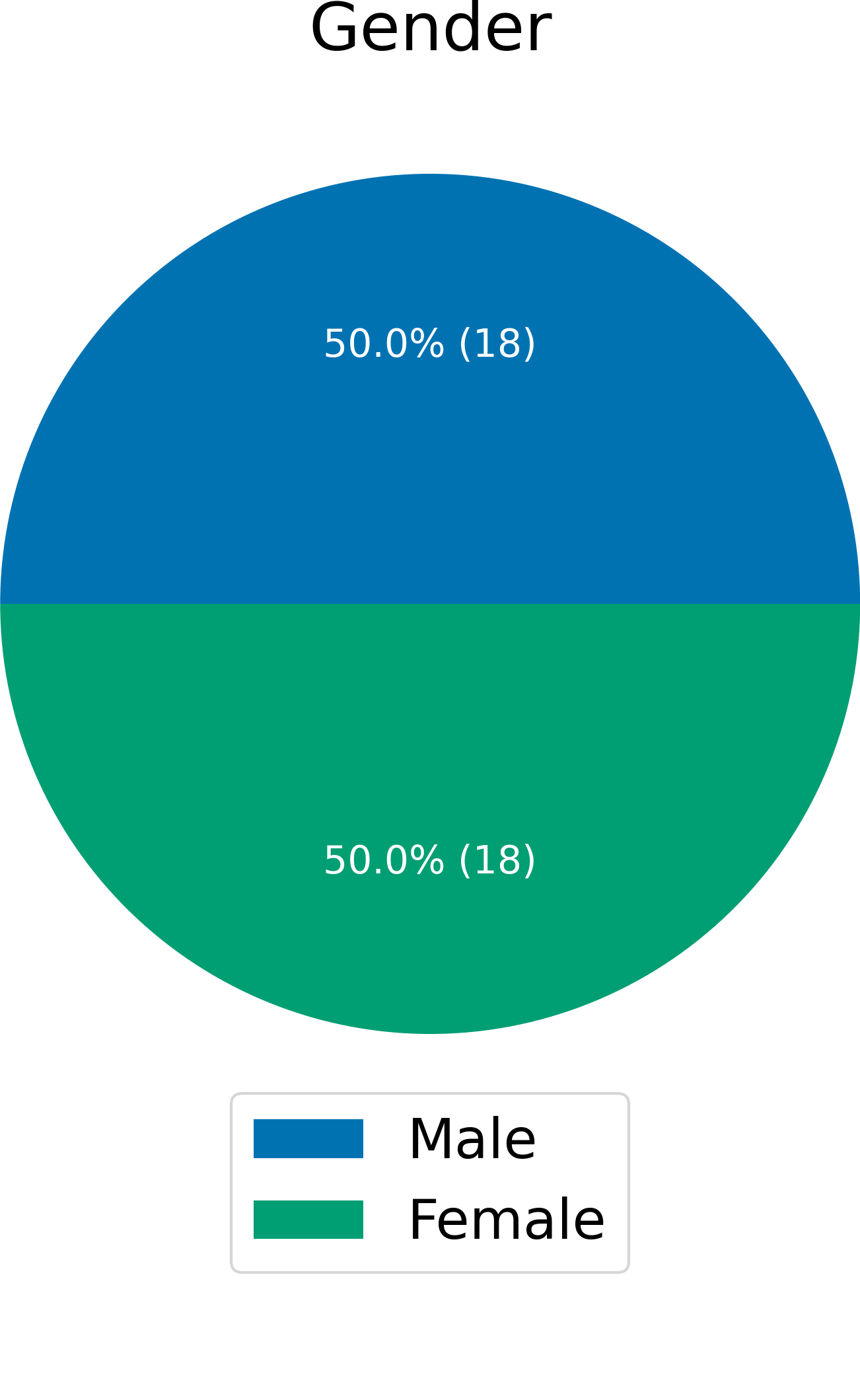}
     \end{subfigure}
     \hfill
     \begin{subfigure}{0.2\textwidth}
         \centering
         \includegraphics[width=\textwidth,valign=t]{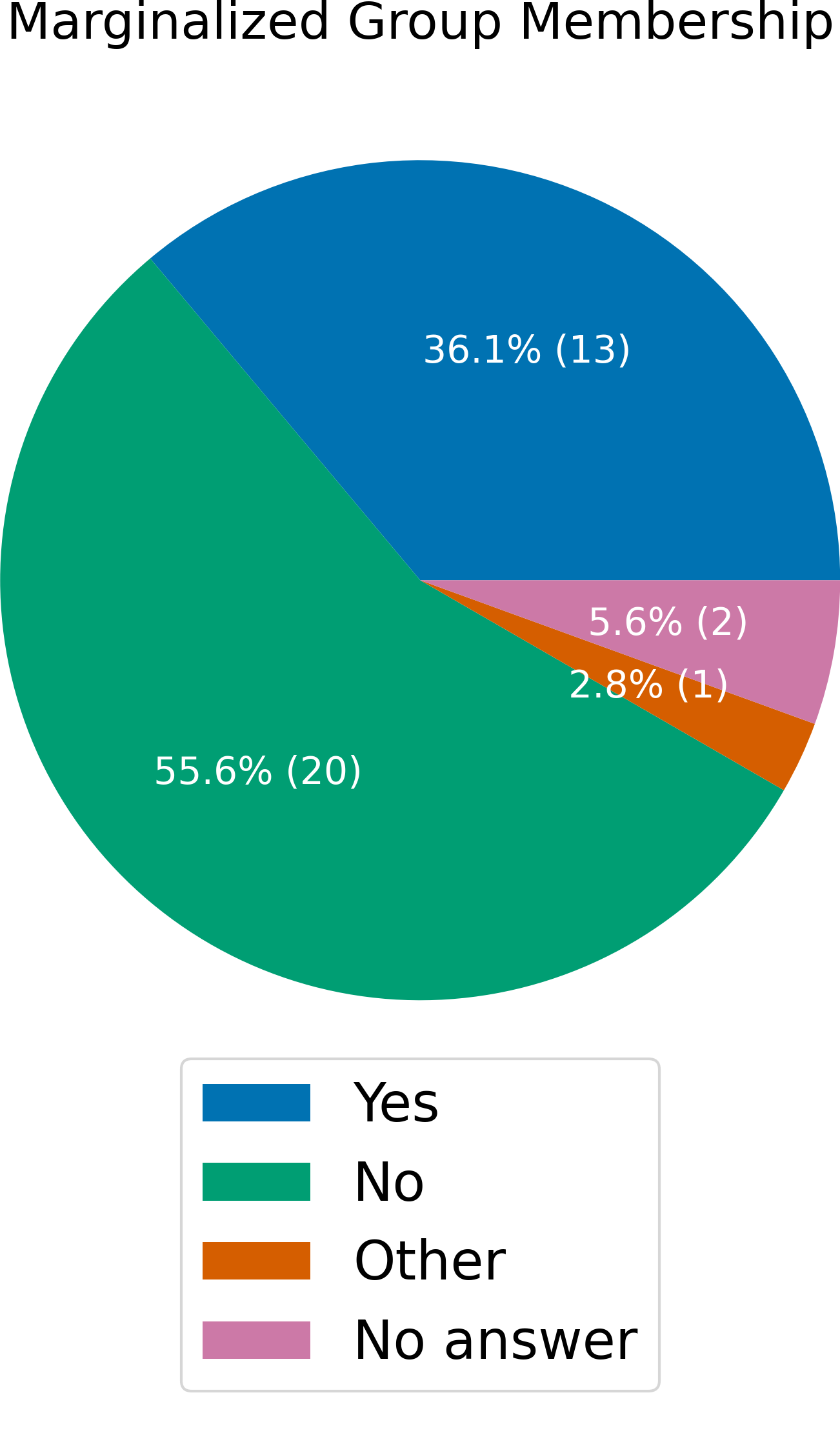}
     \end{subfigure}
        \caption{Participants' demographics and backgrounds. The sample appear relatively diverse (e.g. across gender, political leanings, and marginalized group membership) while it is highly homogeneous in other respects (e.g. most students have a STEM background, and belong to the same age and racial groups).}
        \label{fig:participant-distribution}
\end{figure}

\section{Statistical Tests}\label{app:tests}

Using the pre-activity and post-activity questionnaire responses pertaining to different RAI values as per \citep{jakesch2022different}, we generated a two-way contingency table as shown in Table~\ref{tab:cont-table}. Note that this is a matrix form of the changes represented in Figure~\ref{fig:changes}. Given this categorical data, we then performed two statistical tests of marginal homogeneity: a Bhapkar test \cite{bhapkar1966note} and a Stuart-Maxwell test \cite{stuart1955test,maxwell1970comparing}. More specifically, we conducted these tests to investigate whether we could reject the null hypothesis: our classroom activity did not induce a change in value prioritization and we observed these differences by chance. Even though as per \citet{keefe1982relationship}, the Bhapkar test is generally preferred as it is more powerful, we report results with both tests here.

\begin{table}[h]
    \centering
    \begin{tabular}{c|cccccc}
      & 0 & 1 & 2 & 3 & 4 & 5 \\
      \hline
    0 & 2 & 1 & 0 & 0 & 0 & 0 \\ 
    1 & 2 & 4 & 0 & 0 & 3 & 0 \\ 
    2 & 0 & 0 & 0 & 0 & 0 & 0 \\ 
    3 & 0 & 1 & 0 & 2 & 0 & 0 \\ 
    4 & 0 & 0 & 0 & 0 & 8 & 1 \\ 
    5 & 2 & 0 & 1 & 0 & 1 & 2 
    \end{tabular}
    \caption{Contingency table based on pre-activity and post-activity data from our classroom activity. Rows correspond to pre-activity responses and columns correspond to post-activity responses. Each row or column refers to a different RAI value as outlined in \citet{jakesch2022different}. Category 0 refers to `Accountability \& governance',
 category 1 refers to `Fairness',
 category 2 refers to `Performance \& efficiency',
 category 3 refers to `Privacy',
 category 4 refers to `Safety',
 and category 5 refers to 'Transparency'. This information is also reflected in Figure \ref{fig:changes}.}
    \label{tab:cont-table}
\end{table}

Based on implementations in the DescTools package in R, the Bhapkar test yielded a p-value of 0.2217 and the Stuart-Maxwell test yielded a p-value of 0.34. At significance level $\alpha=0.05$, neither of these results are statistically significant, which in turn means we fail to reject the null. However, we refer to existing literature such as \citet{Vuolo_Uggen_Lageson_2016} that highlights how tests like Stuart-Maxwell require hundreds of paired samples in order to test for conventional levels of significance. Given that our results with only 30 samples still yielded p-values somewhat close 0.05, we argue that this evidences the need for further research with more data for generalization and significance purposes.

\section{Reported AIID Usability Ratings}\label{app:usability}

\begin{figure}[H]
    \centering
    \includegraphics[width=0.3\textwidth]{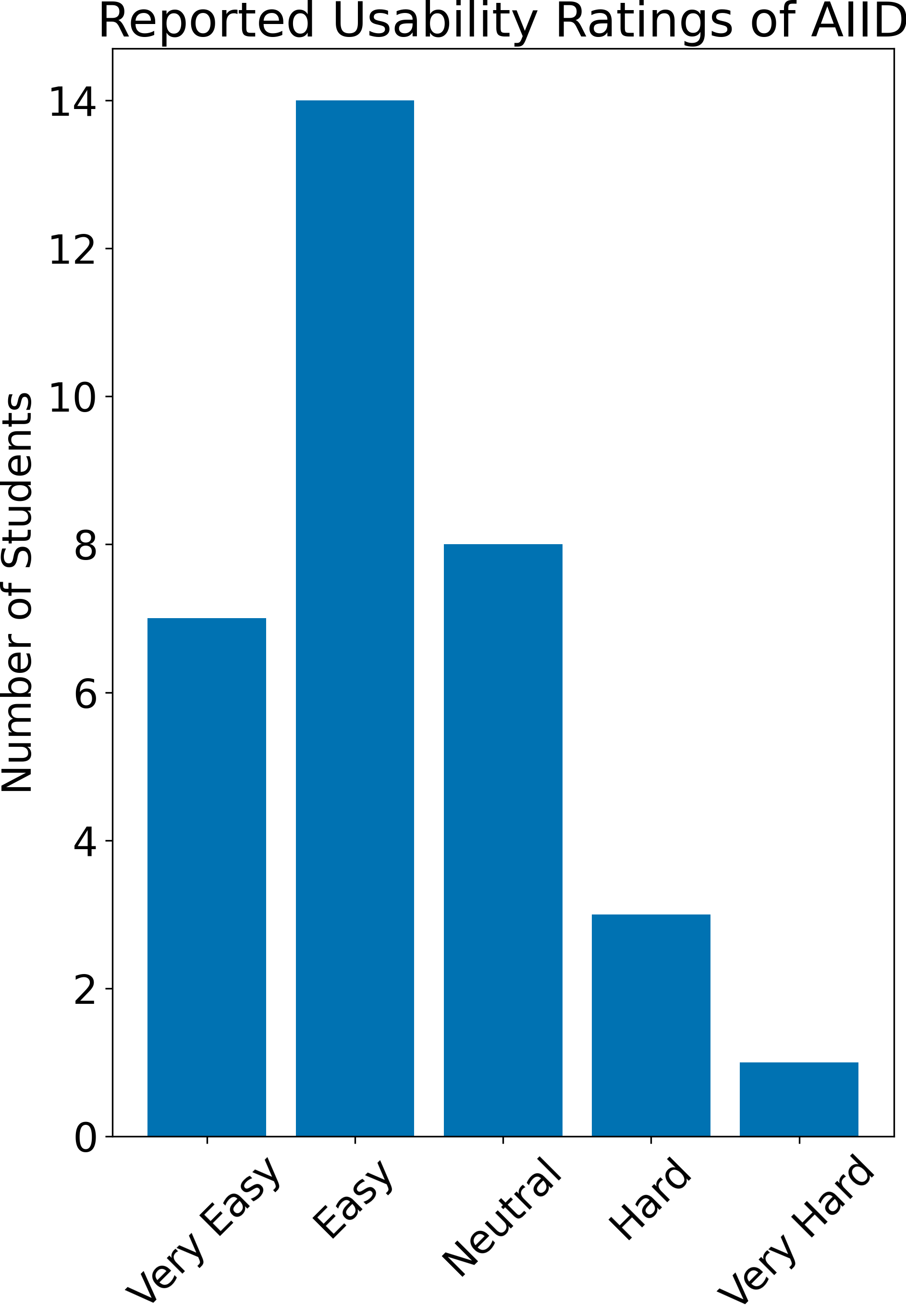}
    \caption{\protect\rule{0ex}{5ex}Student feedback on the usability of the AI Incident Database. While the majority of students reported that the database was ``Easy'' or ``Very Easy'' to use, a nontrivial number of students were ``Neutral'' or stated that it was ``Hard'' or ``Very Hard'' to use. This suggests that while it is already useful, enhancements to the tool would increase its utility.}
    \label{fig:usability-ratings}
\end{figure}

\section{AIID Review Queue}\label{app:queue}
\begin{figure}[H]
    \centering
    \includegraphics[width=\textwidth]{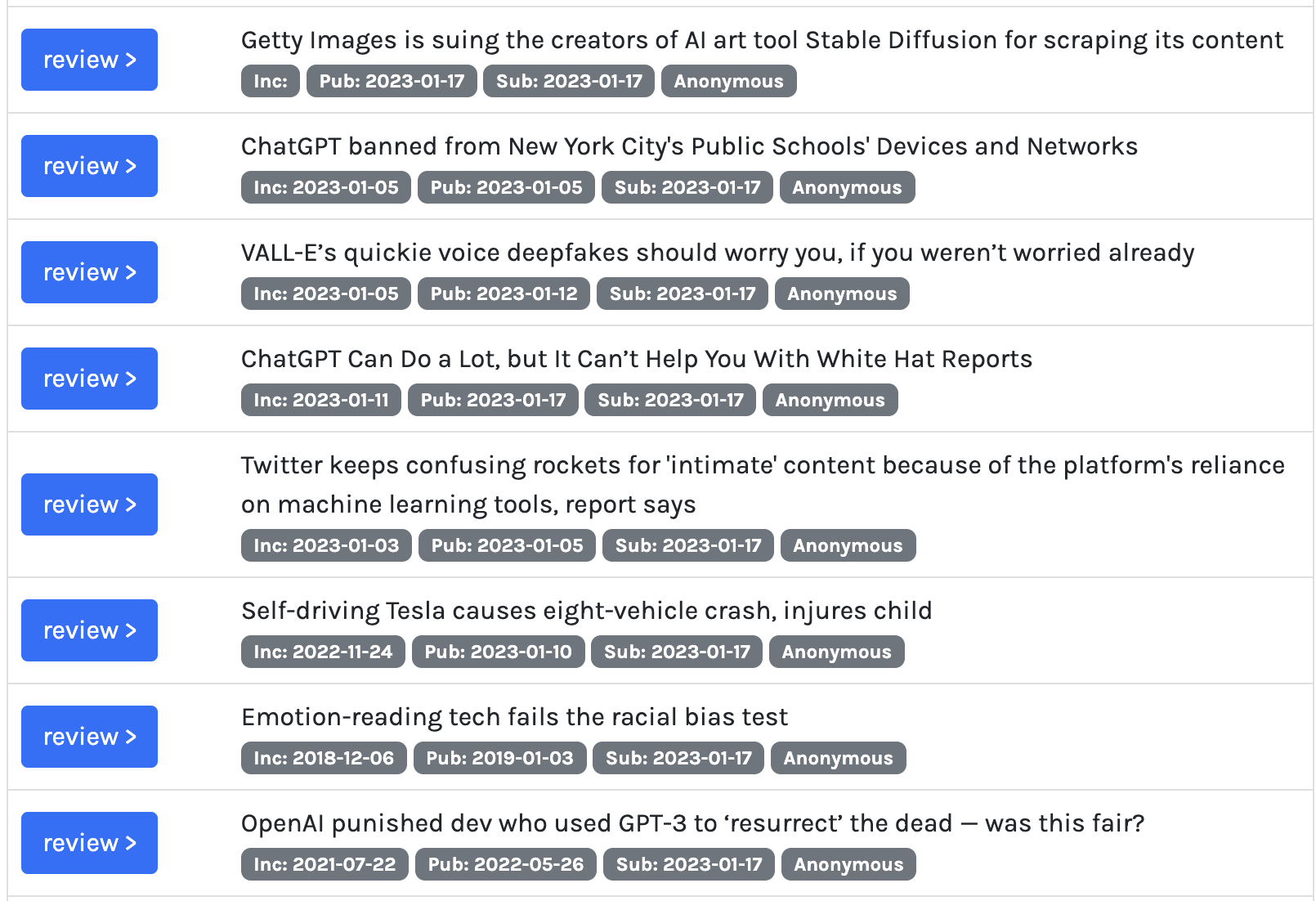}
    \caption{Snapshot of review queue containing reports submitted as part of our activity.}
    \label{fig:queue-snapshot}
\end{figure}      
\end{document}